\documentclass[pre, twocolumn, preprintnumbers, amsmath, amssymb, nofootinbib, floatfix, superscriptaddress]{revtex4}

\usepackage{graphicx, bm, xcolor, bbold}
\usepackage{enumerate}
\usepackage[normalem]{ulem} 
\usepackage[breaklinks]{hyperref}

\makeatletter
\def\graphicscale{\twocolumn@sw{0.3}{0.4}}
\def\graphicthreescale{\twocolumn@sw{0.3}{0.4}}

\begin{document}

\title{Stacked quantum Ising systems and quantum Ashkin-Teller model}

\author{Davide Rossini}
\affiliation{Dipartimento di Fisica dell'Universit\`a di Pisa
  and INFN, Largo Pontecorvo 3, I-56127 Pisa, Italy}

\author{Ettore Vicari}
\affiliation{Dipartimento di Fisica dell'Universit\`a di Pisa,
  Largo Pontecorvo 3, I-56127 Pisa, Italy}

\date{\today}

\begin{abstract}
  We analyze the quantum states of an isolated composite system
  consisting of two stacked quantum Ising (SQI) subsystems, coupled by
  a local Hamiltonian term that preserves the ${\mathbb Z}_2$ symmetry
  of each subsystem. The coupling strength is controlled by an
  intercoupling parameter $w$, with $w=0$ corresponding to decoupled
  quantum Ising systems.  We focus on the quantum correlations of one
  of the two SQI subsystems, ${\cal S}$, in the ground state of the
  global system, and study their dependence on both the state of the
  weakly-coupled complementary part ${\cal E}$ and the intercoupling
  strength.  We concentrate on regimes in which ${\cal S}$ develops
  critical long-range correlations. The most interesting physical
  scenario arises when both SQI subsystems are critical. In
  particular, for identical SQI subsystems, the global system is
  equivalent to the quantum Ashkin-Teller model, characterized by an
  additional ${\mathbb Z}_2$ interchange symmetry between the two
  subsystem operators. In this limit, one-dimensional SQI systems
  exhibit a peculiar critical line along which the length-scale
  critical exponent $\nu$ varies continuously with $w$, while
  two-dimensional systems develop quantum multicritical behaviors
  characterized by an effective enlargement of the symmetry of the
  critical modes, from the actual ${\mathbb Z}_2\oplus {\mathbb Z}_2$
  symmetry to a continuous O(2) symmetry.
\end{abstract}

\maketitle

\section{Introduction}
\label{intro}

Any subsystem of an isolated quantum system made up of multiple
components, such as quantum particles in a gas or quantum spins on a
lattice, can be seen as an open system in contact with an effective
bath (i.e., the remainder of the global system).  In this respect, one
may study the nonunitary quantum dynamics of the subsystem as if it
were subject to the interaction with some environment, while the
global system evolves unitarily~\cite{Zurek-82, Zurek-03}.  Several
paradigmatic and relatively simple composite models have been
considered in the literature. We mention, for example, the so-called
{\em central-spin} models, where one or few qubits are globally
coupled to an environmental many-body system (see, e.g.,
Refs.~\cite{Gaudin-76, PS-00, CPZ-05, QSLZS-06, RCGMF-07, CFP-07,
  YZL-07, CP-08, Zurek-09, BESSS-10, DQZ-11, WL-12, NDD-12, SND-16,
  JH-17, V-18, FCV-19, RV-19, HGMPM-19, LZZ-19, LSSWY-21, YZWS-25}),
models in which a single spin is locally coupled to a many-body
environment~\cite{RCGMF-07, LHMC-08, VPM-15, YZ-20}, {\em sunburst}
spin models, where sets of isolated qubits are locally coupled to a
many-body system~\cite{FRV-22,FRV-22-2,MS-23,FT-23,MS-24}, {\em stacked}
quantum many-body systems~\cite{BYZB-20, BBSD-21, MFPH-23, FPV-23},
etc.  The quantum correlations and decoherence properties of the
subsystems turn out to depend crucially on the global features of
their state, whether a given subsystem is in an ordered or a
disordered quantum phase, or it is close to a quantum critical point,
where large-scale critical correlations
develop~\cite{SGCS-97,Sachdev-book,RV-21}.

\begin{figure}[!b]
  \includegraphics[width=0.95\columnwidth]{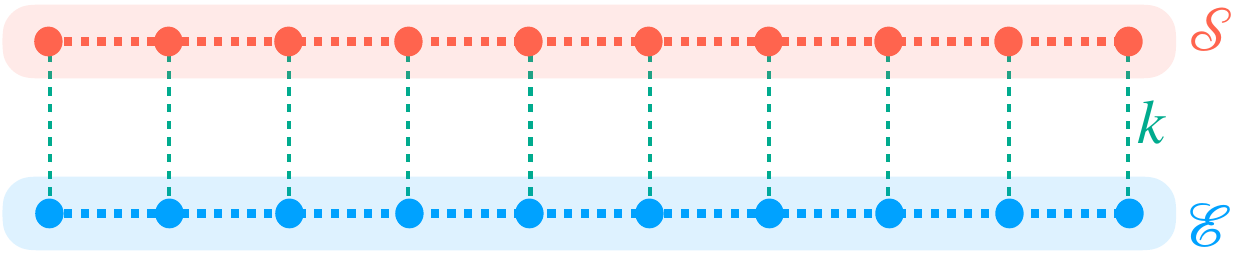}
  \caption{Sketch of a system made of two SQI chains, weakly coupled
    by local and homogeneous interactions controlled by a single,
    generic, parameter $k$.  One of the chains represents the
    subsystem ${\cal S}$, while the other one plays the role of the
    environment ${\cal E}$.}
  \label{sketchsystem}
\end{figure}

To further investigate the quantum dynamics of composite systems, we
consider stacked quantum Ising (SQI) systems, locally and
homogeneously coupled, as sketched in Fig.~\ref{sketchsystem} for a
one-dimensional (1D) setup, as theoretical laboratories.  One of the
two subsystems represents the part ${\cal S}$ under observation, while
the other plays the role of the environment ${\cal E}$. Specifically,
we study the case in which the subsystems are locally coupled by
Hamiltonian terms that do not break the ${\mathbb Z}_2$ symmetries of
each SQI subsystem, such as a local coupling proportional to the
product of their transverse spin operators. The Hamiltonian parameters
of the two subsystems may differ, so that weakly coupled ${\cal
  S}$ and ${\cal E}$ can effectively lie in different quantum phases.
We analyze the quantum correlations of ${\cal S}$ within the ground
state of the global system ${\cal S}\oplus {\cal E}$, focusing on
regimes in which ${\cal S}$ develops critical long-range correlations,
and study their dependence on the state of the weakly-coupled
complementary part ${\cal E}$ and on the intercoupling strength.  By
means of renormalization-group (RG) arguments and extensive numerical
simulations based on the density-matrix RG (DMRG)
algorithm~\cite{Schollwock-05}, we investigate how a weak interaction
between ${\cal S}$ and ${\cal E}$ affects the scaling behaviors of
${\cal S}$, close to quantum critical regimes in the limit of zero
temperature.  We mostly analyze 1D SQI systems, and eventually extend
the discussion to higher-dimensional models, in particular to
two-dimensional (2D) SQI systems.

It is worth noting that one may also consider SQI systems with
different types of local intercoupling, such as that considered in
Ref.~\cite{FPV-23}, where the interaction term between ${\cal S}$ and
${\cal E}$ is proportional to the product of the longitudinal spin
variables. In that case, the global symmetry is altered: the
intercoupling breaks the individual ${\mathbb Z}_2$ symmetries of the
Ising subsystems, leaving only a global residual ${\mathbb Z}_2$
symmetry. Substantial differences emerge when comparing the critical
behaviors of SQI models that preserve or break the original symmetries
of the decoupled subsystems.  Nevertheless, in both cases the critical
behavior developed within ${\cal S}$ depends crucially on the state of
${\cal E}$, whether it is critical or far from criticality.

The paper is organized as follows.  In Sec.~\ref{2Ismod} we introduce
the 1D SQI model studied in this work and define some observables
defined within the subsystem ${\cal S}$. In Sec.~\ref{smallw} we
discuss the critical behavior of ${\cal S}$, when the environment
${\cal E}$ is weakly coupled and far from criticality, being
effectively in a disordered or ordered state.  In Sec.~\ref{atchain}
we focus on the most interesting case in which both subsystems develop
critical correlations, and present a numerical analysis of the
finite-size scaling (FSS) behavior in the limit of equal SQI
subsystems, which becomes equivalent to the quantum Ashkin-Teller
(QAT) model. In Sec.~\ref{d2stacked} we extend our discussion to
higher dimensions, and in particular to 2D SQI models, arguing the
emergence of a peculiar multicritical behavior characterized by an
effective extension of the global symmetry.  Finally, in
Sec.~\ref{conclu} we summarize and draw our conclusions.  In
App.~\ref{app:DMRG} we provide some details on our DMRG simulations
and on their accuracy, while in App.~\ref{isichain} we report some
results for the behavior of critical two-point correlation functions
of 1D quantum Ising chains, obtained by conformal field theory (CFT).

\section{Stacked quantum Ising chains}
\label{2Ismod}

\subsection{The model}
\label{model}

We consider 1D SQI models coupled by local interactions which preserve
their ${\mathbb Z}_2$ symmetry~\cite{SGCS-97,Sachdev-book,RV-21},
as described by the Hamiltonian
\begin{equation}
  \hat H = \hat H_\sigma + \hat H_\tau + \hat H_w,
  \label{twoisiham}
\end{equation}
where
\begin{subequations}
  \label{twoisiham1}
\begin{align}
  & \hat H_\sigma = - J
  \sum_{x=-\ell}^{\ell-1} \hat\sigma_{x}^{(1)}
  \hat\sigma_{x+1}^{(1)} - g \sum_{x=-\ell}^\ell
  \hat\sigma_{x}^{(3)} , \label{Hsig}\\
  & \hat H_\tau = - J_e
  \sum_{x=-\ell}^{\ell-1} \hat\tau_{x}^{(1)}
  \hat\tau_{x+1}^{(1)} - g_e \sum_{x=-\ell}^\ell
  \hat\tau_{x}^{(3)} , \label{Htau}\\
  & \hat H_w = - w \! \left[ 
  \sum_{x=-\ell}^{\ell-1} \hat\sigma_{x}^{(1)}
  \hat\sigma_{x+1}^{(1)} \hat\tau_{x}^{(1)}
  \hat\tau_{x+1}^{(1)} + \! \sum_{x=-\ell}^\ell
  \hat\sigma_{x}^{(3)} \hat\tau_{x}^{(3)}\right] \!. \label{Hsigtau}
\end{align}
\end{subequations}
Here, $\hat\sigma_{x}^{(k)}$ and $\hat\tau_x^{(k)}$ are two sets of
Pauli matrices ($k=1,2,3$) on each site $x$, with $-\ell \le x \le \ell$.
We generally consider systems in which each of the two chains has open
boundary conditions (OBC) and is composed of $L = 2\ell + 1$ sites.
In the following, we also set $\hslash=1$.

The model~\eqref{twoisiham}-\eqref{twoisiham1} has a global
${\mathbb Z}_2\oplus{\mathbb Z}_2$ symmetry for generic values of
the parameters $J, g, J_e, g_e$, and $w$, combining the
two independent parity symmetries
\begin{subequations}
  \label{paritysym}
  \begin{align}
    &\hat\sigma_{x}^{(1)}
    \rightarrow - \hat\sigma_{x}^{(1)},\qquad
    \hat\sigma_{x}^{(3)} \rightarrow \hat\sigma_x^{(3)},\\
    &\hat\tau_{x}^{(1)}
    \rightarrow - \hat\tau_{x}^{(1)},\qquad
    \hat\tau_{x}^{(3)} \rightarrow \hat\tau_x^{(3)}.
  \end{align}
\end{subequations}
When $J=J_e$ and $g=g_e$, the global symmetry
extends to a further interchange ${\mathbb Z}_2$ symmetry
\begin{equation}
  \hat{\bm \sigma}_{x} \leftrightarrow \hat{\bm \tau}_{x} .
  \label{intersym}
\end{equation}      
In the latter case, we recover the 1D QAT
model~\cite{KNK-81,IS-84,AF-84,Fradkin-84,Shankar-85,GR-87,ABB-88},
whose critical properties are related to those of the 2D classical
Ashkin-Teller (CAT) model~\cite{AT-43,WL-74} by a quantum-to-classical
mapping.

When the interaction between the SQI chains vanishes (i.e., when
$w=0$), one recovers two decoupled quantum Ising chains, with
Hamiltonians $\hat H_\sigma$ and $\hat H_\tau$, respectively. The
single chain undergoes a zero-temperature continuous quantum transition
at a critical value $g_{\cal I}$ of the transverse field,
\begin{equation}
  g_{\cal I} = J,
  \label{gcgi}
\end{equation}
see, e.g., Refs.~\cite{Sachdev-book,RV-21,DACDRS-book}.  The
corresponding quantum critical behavior belongs to the 2D Ising
universality class.  The deviation $r \equiv g-g_{\cal I}$ represents
the leading RG perturbation which preserves the ${\mathbb Z}_2$
symmetry.  Its RG dimension is $y_r=1/\nu_{\cal I}=1$, so that the
length scale $\xi$ of the critical modes behaves as $\xi\sim
|g-g_{\cal I}|^{-\nu_{\cal I}}$.  The dynamic exponent, controlling
the vanishing of the gap $\Delta\sim \xi^{-z}$ at the transition
point, is $z=1$.  The RG dimension $y_\phi$ of the longitudinal spin
operator $\hat \sigma_x^{(1)}$, associated with the order parameter at
the Ising transition, is given by $y_\phi=\eta/2=1/8$, where
$\eta=1/4$ is the critical exponent characterizing the spatial decay
of the critical correlations of $\hat \sigma_x^{(1)}$.  For any
$g<g_{\cal I}$, the presence of a homogeneous longitudinal field $h$
coupled to the spin operator $\hat\sigma_x^{(1)}$ drives first-order
quantum transitions at $h=0$. In particular, for $|h|\to 0$, the
magnetized states $|+\rangle$ and $|-\rangle$ along the longitudinal
direction are the ground states of the system for $h>0$ and $h<0$,
respectively, giving rise to a discontinuous infinite-volume
longitudinal magnetization $m_0$:~\cite{Pfeuty-70}
\begin{equation}
  \lim_{h\to 0^\pm} \lim_{L\to\infty} M = \pm \, m_0, \qquad m_0 = (1
  - g^2)^{1/8},
  \label{sigmasingexp}
\end{equation}
where $M$ is the ground-state expectation value of the operator
$\hat M =L^{-1} \sum_x \hat\sigma_x^{(1)}$.

In this paper, we present a numerical study of the zero-temperature
phase diagram and the critical properties of the global
model~\eqref{twoisiham}-\eqref{twoisiham1}, based on extensive DMRG
simulations~\cite{Schollwock-05} for systems with up to $L=65$
sites. Further details on the choice of the parameters for the
numerical algorithm and on the accuracy of our simulations are
provided in App.~\ref{app:DMRG}.  In our study we set $J_e=J=1$
without loss of generality (unless some specific limits are
considered), and discuss quantum correlations within the ground state
of the global model, in particular within ${\cal S}$, when varying the
Hamiltonian parameters $g,\,g_e$, and $w$.

It is worth mentioning that other types of couplings between the two
SQI chains can also be considered, which break, or partially break,
the ${\mathbb Z}_2\oplus {\mathbb Z}_2$ symmetry of the decoupled
quantum Ising systems. For example, one may consider the same
Hamiltonian model~\eqref{twoisiham}, in which the local interaction
term $\hat H_w$ of Eq.~\eqref{Hsigtau} is replaced by
\begin{equation}
  \hat H_\kappa =
  -\kappa \sum_x \hat{\sigma}_{x}^{(1)} \hat{\tau}_{x}^{(1)},
  \label{Hkint}
\end{equation}
which breaks the ${\mathbb Z}_2$ symmetries of the Ising chains,
leaving a residual combined ${\mathbb Z}_2$ symmetry when one
simultaneously changes $\hat\sigma_{x}^{(1)} \rightarrow -
\hat\sigma_{x}^{(1)}$ and $\hat\tau_{x}^{(1)} \rightarrow -
\hat\tau_{x}^{(1)}$.  A study of the effects of the interaction
$\hat H_\kappa$ on SQI systems has been reported in Ref.~\cite{FPV-23}.
As we shall see below, different scenarios emerge when the interaction
between the stacked systems preserves the ${\mathbb Z}_2$ symmetries
of the subsystems, such as the one described by the Hamiltonian term
$\hat H_w$ in Eq.~\eqref{Hsigtau}.

\subsection{Observables}
\label{obs}

To study the equilibrium properties of the subsystem ${\cal S}$ when
the global system is in the ground state $|\Psi_0\rangle$, considering
the other Ising chain as the environment ${\cal E}$, one may introduce
its reduced density matrix
\begin{equation}
  \hat \rho_{\cal S} = {\rm Tr}_{\cal E} \, \big[|\Psi_0\rangle
    \langle \Psi_0|\big],
  \label{rhoIA}
\end{equation}
where ${\rm Tr}_{\cal E} [\,\cdot\,]$ denotes the partial trace over
the Hilbert space associated with the subsystem ${\cal E}$.

Due to the global ${\mathbb Z}_2$ symmetry, the expectation value of
$\hat{\sigma}_x^{(1)}$ vanishes, i.e.,
${\rm Tr} \big[ \hat \rho_{\cal S}\, \hat{\sigma}_x^{(1)} \big] = 0$.
Therefore we consider its two-point correlation function
\begin{equation}
  G(x,y) \equiv {\rm Tr} \big[ \hat \rho_{\cal S} \, \hat{\sigma}_x^{(1)}
    \hat{\sigma}_y^{(1)} \big] .
  \label{gxy}
\end{equation}
We recall that translation invariance is not preserved in systems with
OBC, thus $G(x,y)$ depends on both spatial positions $x$ and $y$.
Since we consider odd values of $L=2\ell+1$, and choose coordinates
such that $-\ell \le x \le \ell$, we identify a central site, $x_0=0$.

In the following, we mostly focus on the correlations
\begin{equation}
  G_0(x) \equiv G(x_0,x)
  \label{g0def}
\end{equation}
between the central point and the other sites.  We also consider the
susceptibility $\chi_0$ and the second-moment correlation length
$\xi_0^2$ associated with the two-point function $G_0(x)$, defined as
\begin{equation}
  \chi_0 = \sum_x G_0(x),\qquad 
  \xi_0^2 = {1\over 2  \chi_0} \sum_x x^2 G_0(x).
  \label{chixidef}
\end{equation}
Within the subsystem ${\cal S}$, one may also address the correlations
of the transverse operators $\hat\sigma_x^{(3)}$, defined as
\begin{equation}
  F(x,y) \equiv
  {\rm Tr} \big[ \hat \rho_{\cal S} \, \hat \sigma_x^{(3)}
    \hat \sigma_y^{(3)} \big] -
  {\rm Tr} \big[ \hat \rho_{\cal S} \, \hat \sigma_x^{(3)}\big] \:
  {\rm Tr} \big[ \hat \rho_{\cal S} \, \hat \sigma_y^{(3)} \big],
  \label{fxy}
\end{equation}
and the corresponding $F_0(x)\equiv F(x_0,x)$.

\section{Scaling behaviors for weakly coupled Ising chains}
\label{smallw}

We now discuss the scaling behaviors of weakly interacting Ising
chains, i.e., for small values of $w$, when the subsystem ${\cal S}$
is close to criticality ($g\approx g_{\cal I}=1$).  As shown in
Ref.~\cite{FPV-23}, when the intercoupling is driven by $\hat
H_\kappa$ [cf. Eq.~\eqref{Hkint}], the critical dependence on the
coupling parameter $\kappa$ strongly depends on the phase of the
environment chain, i.e., whether its parameter $g_e$ is larger, equal,
or smaller than $g_{\cal I}$.  Here, we instead study the critical
scenarios emerging in ${\cal S}$ when the interaction is driven by the
symmetry-preserving Hamiltonian term $\hat H_w$ of
Eq.~\eqref{Hsigtau}.  We anticipate that, while a crucial dependence
on the environmental state persists, substantial differences arise
compared to the scaling behavior found for the symmetry-breaking
interaction $\hat H_\kappa$.

In this section we focus on situations in which the environment
${\cal E}$ is far from criticality, meaning that $g_e$ is sufficiently
different from the Ising critical value $g_e=g_{\cal I}=1$.  The most
interesting case, in which ${\cal E}$ also presents critical
correlations, will be discussed in Sec.~\ref{atchain}.

\subsection{Ising transition lines}
\label{isitralines}

In the following we argue that, for both cases $g_e>g_{\cal I}$ and
$g_e<g_{\cal I}$, corresponding to a disordered or an ordered
environment ${\cal E}$, a weak interaction $\hat H_w$ only gives rise to
a shift of the critical value of the Ising transition,
from $g=g_{\cal I}$ for $w=0$ to
\begin{equation}
  g_c(w,g_e) \approx g_{\cal I} + c(g_e)\,w
  \label{gcw}
\end{equation}
for sufficiently small values of $w$, and the critical behavior of the
correlation functions in ${\cal S}$ remains within the 2D Ising
universality class.

The behavior (\ref{gcw}) can be derived using RG arguments.  Since the
environment chain is supposed to be noncritical and the expectation
value of $\hat \tau_x^{(3)}$ is nonzero for any value of $g_e$ (see,
e.g., Refs.~\cite{Pfeuty-70,CPV-15}), the effect of the intercoupling
$\hat H_w$ should effectively correspond to a further $w$-dependent
term proportional to $\hat \sigma_x^{(3)}$ in the Hamiltonian $\hat
H_\sigma$ associated with the subsystem ${\cal S}$
[cf.~Eq.~\eqref{Hsig}].  Therefore, the quantum Ising transition of
the isolated ${\cal S}$ should only get shifted to a $w$-dependent
critical value $g_c(w,g_e)$.  In other words, the effect of a small
coupling $w$ can be effectively taken into account by adding an
analytical dependence to the even scaling field $u_r$ associated with
the quantum Ising transition, i.e.,
\begin{equation}
  u_r(g,w,g_e) \approx r - c(g_e) \, w,\qquad r = g - g_{\cal I},
  \label{ug}
\end{equation}
where the nonuniversal constant $c$ generally depends on the
environment coupling $g_e$ (we also fix an arbitrary normalization
requiring $u_r\approx r$ for $w = 0$).  Therefore the singular part of
the free-energy density in the zero-temperature and FSS limit is
expected to scale as~\cite{CPV-14,RV-21}
\begin{equation}
  F_{\rm sing}(g,w,L) \approx L^{-(d+z)}{\cal F}(u_r L^{y_r}),
  \label{freeendis}
\end{equation}
where $y_r=1$ is the inverse length-scale critical exponent of the 2D
Ising universality class.  Since the critical point must correspond to
the vanishing of the scaling field $u_r$, we obtain the linear
relation (\ref{ug}) for the critical line at finite small values of
$w$ and for any fixed value of $g_e$ far from the Ising critical value
$g_e\approx g_{\cal I}$.

\subsection{FSS along the Ising transition lines}
\label{fssisi}

To check the prediction~\eqref{gcw} for the $w$-dependence of the
transition lines and the 2D Ising universality class of the critical
behaviors along these transition lines, we present a numerical
analysis based on DMRG computations and their matching with the
expected FSS behaviors at a quantum Ising transition~\cite{CPV-14,
  RV-21}.  We focus on the correlation functions within the subsystem
${\cal S}$ and study the effects of a small intercoupling with the
environment ${\cal E}$ (i.e., for small $w$), when the Hamiltonian
parameter of ${\cal E}$ is far from the critical value $g_{\cal I}$ of
the isolated subsystem. Namely, we consider $g_e=2$ and $g_e=0.5$.

Assuming that the critical behaviors along the transition lines for
small $w$ belong to the 2D Ising universality class, the asymptotic
FSS behavior of the two-point functions of the longitudinal and
transverse spin operators [cf. Eqs.~(\ref{gxy}) and (\ref{fxy})], is
expected to be given by~\cite{CPV-14,RV-21}
\begin{align}
  & G(x_1,x_2) \approx L^{-2y_\phi} {\cal G}(X_1,X_2,u_rL^{y_r}),
  \label{gxysca} \\
  & F(x_1,x_2) \approx L^{-2y_e} {\cal F}(X_1,X_2,u_rL^{y_r}),
  \label{fxysca}
\end{align}
where
\begin{equation}
  X_i = {x_i\over 2\ell},\qquad L=2\ell + 1
  \label{Xidef}
\end{equation}
    [for convenience we rescale distances using $2\ell=L-1$, the
      difference in the asymptotic FSS limit only giving rise to
      $O(L^{-1})$ corrections], $u_r$ is the scaling field given in
Eq.~\eqref{ug}, $y_\phi$ and $y_e$ are the RG dimensions of the
order-parameter field and energy operators, given respectively by
$y_\phi = (d+z-2+\eta)/2=1/8$ and $y_e = d + z - y_r = 2 - y_r=1$.
The scaling functions ${\cal G}$ and ${\cal F}$ are universal with
respect to microscopic details of the model, but they depend on the
boundary conditions. CFT allows us to determine them (see
App.~\ref{isichain} for details).

The ratio between the correlation length $\xi_0$,
cf.~Eq.~(\ref{chixidef}), and the size of the chain,
\begin{equation}
  R_\xi = \xi_0/L,
  \label{rxidef}
\end{equation}
is an RG invariant quantity. In the FSS limit, it scales
as~\cite{CPV-14,RV-21}
\begin{equation}
  R_\xi(g,w,L) = {\cal R}_\xi(u_r L^{y_r}) + O(L^{-\zeta}) + O(L^{-1}),
  \label{rxisca}
\end{equation}
where $u_r$ is given in Eq.~(\ref{ug}), ${\cal R}_\xi$ is a universal
function apart from a rescaling of its argument, and the leading
scaling corrections are controlled by the exponent
$\zeta=2-z-\eta=3/4$.  In particular, at the critical point $g=g_c$,
thus $u_r=0$, $R_\xi$ behaves as~\cite{CPV-14,RV-21}
\begin{equation}
  R_\xi(g_c,w,L) = R_\xi^* + a_\zeta L^{-\zeta} + a_{1} L^{-1}
  + a_\omega L^{-\omega} + \ldots
  \label{rxisca2}
\end{equation}
where $R_\xi^*={\cal R}_\xi(0)$ is an universal value which depends on
the boundary conditions. For OBC, the critical value associated with
the 2D Ising universality class is $R_\xi^* \approx 0.159622$ (see
App.~\ref{isichain}).  There are various sources of power-law
suppressed scaling corrections.  The $O(L^{-\zeta})$ scaling
correction arises from the background analytical term of the
susceptibility at the denominator of the second-moment correlation
length~\cite{CPV-14,RV-21}.  The $O(L^{-1})$ correction is instead
related to the boundary corrections associated with the OBC.  Finally,
the $O(L^{-\omega})$ scaling correction with $\omega=2$ is related to
the leading irrelevant RG perturbation at the 2D Ising fixed
point~\cite{PV-02,CHPV-02}.  Note also that, since $R_\xi$ is
generally a monotonic function of $u_r$, the FSS behavior
(\ref{rxisca}) implies that the curves of $R_\xi$ for different
lattice sizes $L$ and $L_2>L$ cross each other, at a value of $g$ that
approaches $g_c$ with $O(L^{-1/\nu-\zeta})$ corrections, thus
$O(L^{-7/4})$ for Ising-like transitions, and the corresponding value
of $R_\xi$ approaches $R_\xi^*$ with corrections that decay as
$L^{-\zeta}$.

Other RG invariant quantities are obtained by taking the ratios of the
correlation function $G$ at different distances, such as
\begin{equation}
  R_G(X_1,X_2) = {G_0(x_1=2 \ell X_1)\over G_0(x_2=2\ell X_2)},
  \label{rgdef}
\end{equation}
which are expected to scale similarly to $R_\xi$, i.e.,
\begin{equation}
  R_G(X_1,X_2,g,w,L) = {\cal R}_G(X_1,X_2,u_r L^{y_r}) + O(L^{-1})
  \label{rgsca}
\end{equation}
where ${\cal R}_G$ is a function that is universal apart from a
rescaling of its argument $u_r L^{y_r}$~\cite{PV-02,CPV-14}.  In
particular, at the critical point, $R_\xi$ behaves as~\cite{CPV-14}
\begin{equation}
  R_G(X_1,X_2) = R_G(X_1,X_2)^* + O(L^{-1}).
  \label{rgGsca2}
\end{equation}
Using CFT, one can compute $R_G(X_1,X_2)^*$, for example
$R_G^*(1/8,1/4)\approx 1.358609$ (see App.~\ref{isichain}).  In the
above formulas for $R_G$, the leading scaling corrections are expected
to arise from the boundaries, thus they are $O(L^{-1})$ for OBC. Note
that the ratios $R_G$, unlike $R_\xi$, are not affected by the
analytical $O(L^{-\zeta})$ contributions at the critical point.

\subsection{Numerical results}
\label{numres}

\begin{figure}[!t]
  \includegraphics[width=0.47\textwidth]{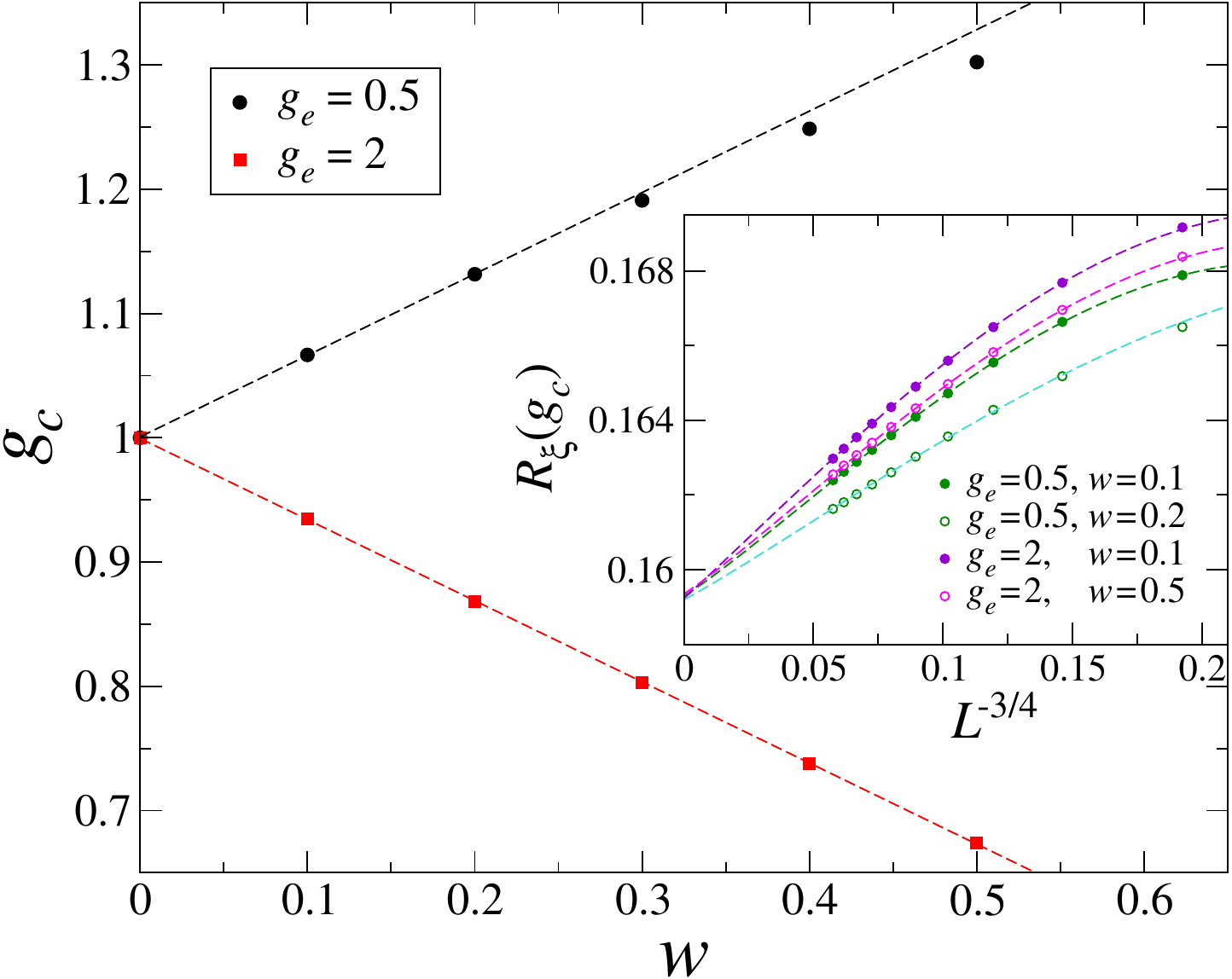}
  \caption{The critical value $g_c$ of the Ising transition as a
    function of $w$ for a model of weakly coupled Ising chains,
    described by the Hamiltonian~\eqref{twoisiham}-\eqref{twoisiham1}.
    The two data sets are for $g_e=0.5$, black dots, and $g_e=2$, red
    squares (the uncertainty on the estimates of $g_c$ are smaller than
    the symbols). The dashed lines show linear fits of the critical
    points for $w\lesssim 0.2$ to $g_c(w,g_e) = g_{\cal I} + c(g_e) w$,
    which is the expected behavior for sufficiently small
    values of $w$ (see text). The inset shows $R_\xi$ vs $L^{-3/4}$ at
    the crossing point between data for sizes $L$ and $L+4$, at fixed
    $g_e$ and $w$ (symbols correspond to sizes ranging from $L=9$ to
    $L=45$, in increments of four).  Dashed curves are fits to the
    corresponding numerical data, for $L \geq 13$, according to the
    predicted behavior~\eqref{rxisca2}.}
\label{gcwge}
\end{figure}

We now present a numerical analysis for $g_e=2$ and $g_e=0.5$,
at small values of $w$, for values of $g$ around $g_{\cal I}=1$.
As we shall see, the results nicely match the FSS
behaviors discussed above, confirming that the critical behavior of
the transition lines originating from $g=1$ and $w=0$ remains
Ising-like.

The $w$-dependence of the critical points $g_c(w,g_e)$ can be
straightforwardly obtained by matching our data (we performed 
simulations up to $L=49$) for the RG invariant quantities $R_\xi$ and
$R_G(X_1,X_2)$ at fixed small $w$ with their expected FSS behavior at
Ising-like transitions [see Eqs.~\eqref{rxisca} and~\eqref{rgsca}].
In particular, the values at the crossing points of $R_\xi$
and $R_G(1/8,1/4)$ for different sizes appear to converge to the
corresponding universal values of the 2D Ising universality class,
namely $R_\xi^* = 0.159622$ and $R_G^*(1/8,1/4)=1.358609$ (see
App.~\ref{isichain}). For example, the inset of Fig.~\ref{gcwge} shows
the crossing points of $R_\xi$ for sizes $L$ and $L+4$ plotted versus
$L^{-3/4}$, corresponding to the expected leading power-law scaling
corrections. They appear to converge to the Ising value with an
accuracy of about $5 \times 10^{-4}$. These results
confirm that the transitions belong to the 2D Ising universality class.

The critical points $g_c(w,g_e)$ can be then straightforwardly
obtained by fitting the $R_\xi$ data to the FSS ansatz~\eqref{rxisca},
using the Ising values $R_\xi^* = 0.159622$ and $\nu=1$, and also
including the expected $O(L^{-3/4})$ scaling corrections. Close to
$g_c$, low-order polynomial approximations of the scaling function
${\cal R}_\xi(u_r L^{1/\nu})$ are sufficient.  Analogous results are
obtained by analyzing the data of the ratios $R_G(X_1,X_2)$.
This procedure allows us to obtain accurately estimates of
$g_c(w,g_e)$, as shown in Fig.~\ref{gcwge}. These results are in good
agreement with the linear $w$-dependence predicted by the RG arguments
reported in Sec.~\ref{isitralines} [cf.~Eq.~\eqref{gcw}].  Note,
however, that deviations from the expected behavior emerge at smaller
values of $w$ for $g_e=0.5$ than for $g_e=2$.

We finally remark that the scaling behavior of the subsystem ${\cal S}$
appears to be the same, when the Hamiltonian parameter $g_e$ of
the environment ${\cal E}$ corresponds to either the paramagnetic
($g_e>1$) or the ferromagnetic ($g_e<1$) phase. This behavior is
markedly different from that observed when the interactions between
the SQI subsystems are driven by the Hamiltonian term~\eqref{Hkint},
which breaks the global symmetry of the decoupled subsystems.  In that
case, the scaling behaviors turn out to substantially differ between
$g_e>1$ and $g_e<1$~\cite{FPV-23}.  As we shall see, the scenario
becomes more complex when both subsystems are critical, i.e., when
$g\approx g_e\approx g_{\cal I}$.

\section{Scaling behaviors in the limit of equal subsystems}
\label{atchain}

We now concentrate on the scenario in which the SQI chains
are critical, with both couplings $g$ and $g_e$ close to the
critical value $g_{\cal I}=1$. To simplify our analysis, we consider
the symmetric case $g=g_e$, corresponding to the 1D QAT model
described by Hamiltonian~\eqref{twoisiham}-\eqref{twoisiham1}
with $J=J_e=1$ and $g=g_e$. The QAT model is symmetric under
the independent ${\mathbb Z}_2$ parity symmetries:
$\hat\sigma_{x}^{(1,2)} \rightarrow -\hat\sigma_{x}^{(1,2)}$
keeping $\hat\sigma_{x}^{(3)}$ unchanged,
$\hat\tau_{x}^{(1,2)} \rightarrow - \hat\tau_{x}^{(1,2)}$ keeping
$\hat\tau_{x}^{(3)}$ unchanged, and the ${\mathbb Z}_2$ interchange
symmetry $\hat{\bm \sigma}_{x} \leftrightarrow \hat{\bm \tau}_{x}$.
In particular, for $w=0$, two identical and decoupled quantum Ising
chains are recovered.

Some studies related to the 1D QAT model have been reported in
Refs.~\cite{KNK-81,IS-84,AF-84,Fradkin-84,Shankar-85,GR-87,BGR-87,Yang-87,
  YZ-87,ABB-88,YHK-94,YK-95,BBBD-15,BBDF-15,LMC-24,LCFO-26}.  In the
following, we focus on the FSS behavior of the 1D QAT model along its
continuous quantum transition line where the length-scale critical
exponent $\nu$ varies continuously.

\subsection{The critical line of the QAT chain}
\label{qatmodel}

The phase diagram of the 1D QAT model shows a peculiar line of
continuous transitions with central charge $c=1$, where the
length-scale critical exponent changes continuously, along the line
$g=g_c=1$ and $-1/\sqrt{2} \le w \le 1$.  Along this transition line,
$w$ represents a marginal parameter, i.e., its RG dimension vanishes
at the corresponding line of fixed points.  At fixed $w$, when varying
$g$ around $g_c$, the divergence of the length scale of the critical
modes is controlled by a continuously varying critical exponent $\nu$,
whose dependence on the Hamiltonian parameter $w$ has been predicted
to be~\cite{KNK-81,ABB-88}
\begin{equation}
  \nu(w) = { 2 \, {\rm arccos}(-w) \over 4\,{\rm arccos}(-w) - \pi},
  \label{nuexp}
\end{equation}
on the basis of a mapping to the exactly solvable six-vertex
model~\cite{Wu-77,Baxter-82}.  Therefore, along the transition line,
the length-scale exponent varies from $\nu=2/3$ for $w=1$
(corresponding to the $q=4$ Potts model) to $\nu\to\infty$
for $w \to -1/\sqrt{2}$ (for example we have $\nu=4/5$ for $w=1/2$,
$\nu=1$ for $w=0$, and $\nu=2$ for $w=-1/2$).  On the other hand,
the critical exponent $\eta=1/4$ (associated with the RG dimension
$y_\phi= \eta/2$ of the order-parameter field) and the dynamic
exponent $z=1$ (associated with the power-law suppression
of the gap at the critical point) remain the same along the line.

This peculiar behavior of the critical exponents, and in particular
the dependence of $\nu$ on $w$, have been confirmed by various
numerical analyses (see, e.g.,
Refs.~\cite{ABB-88,YK-95,BBBD-15,BBDF-15}), with a reasonable accuracy
(the most recent Refs.~\cite{BBBD-15,BBDF-15} mention that their
estimates of $\nu$ have a relative accuracy of 5-10\% sufficiently far
from the boundaries $w=-1/\sqrt{2}$ and $w=1$ of the interval where
continuous transitions occur).  We have checked the formula~\eqref{nuexp}
by FSS analyses of our DMRG data at fixed $w$ and around
$g_c=1$. For this purpose, the optimal observables are provided by the
RG invariant ratios $R_G(X_1,X_2)$, defined in Eq.~\eqref{rgdef},
whose scaling behavior at fixed $w$ is expected to be
\begin{equation}
  R_G(X_1,X_2,g,w,L) \approx {\cal R}_G[X_1,X_2,(g-g_c)L^{1/\nu}].
  \label{rgscaw}
\end{equation}
The ratios $R_G$ are expected to yield more precise
results than $R_\xi$ [cf.~Eq.~\eqref{rxidef}], because their FSS is
not affected by analytic corrections (see the discussion in
Sec.~\ref{fssisi}).

A straightforward analysis of our DMRG data for $R_G(1/8,1/4)$
at fixed $w=-0.5$ and $w=+0.5$, up to $L=65$, by matching
them to the asymptotic scaling Eq.~\eqref{rgscaw}, and in particular
to its linear approximation $R_G^*(1/8,1/4) + a (g-g_c)L^{1/\nu}$,
leads to precise estimates $R_G(1/8,1/4) = 1.2456(1)$ and
$\nu=2.02(2)$ for $w=-0.5$, and $R_G(1/8,1/4) = 1.491(1)$ and
$\nu=0.802(8)$ for $w=+0.5$, with a relative accuracy
of about 1\% (these results can be compared with the
Ising values $R_G(1/8,1/4) = 1.358609$ and $\nu=1$).  These findings
are obtained from fits of the largest lattices in the range
$33 \lesssim L \le 65$; the errors take into account the variation
of the fitted values, when changing the minimum system size included
in the fit and the linear fitting ansatzes using $L$ or $2\ell$, thus
somehow quantifying the effects of the $O(L^{-1})$ scaling corrections.
The outcomes are in excellent agreement with the exact results from
Eq.~\eqref{nuexp} (i.e., $\nu=2$ for $w=-0.5$, and $\nu=0.8$ for
$w=0.5$).  We do not report further details, as they are not
particularly illuminating for the purposes of this paper.  In the
following, we assume the validity of the formula in Eq.~\eqref{nuexp}
for the values of the critical exponent $\nu$.

The phase diagram for values of $w$ outside the range
$w\in [-1/\sqrt{2},1]$ presents other transition lines, where 
quantum critical behaviors are expected to belong to the 2D
Ising and Berezinskii-Kosterlitz-Thouless universality
classes~\cite{KNK-81,IS-84,GR-87,YHK-94}.  In particular, the extremal
critical point $g=1$, $w=1$, corresponding to a quantum $q=4$ state
Potts model, is the starting point of two Ising transition lines for
$w>1$ and $g\neq 1$. We also note that a number of studies of the
phase diagram and critical behavior of the 2D CAT model are reported
in Refs.~\cite{WL-74,DR-79,DK-82,WD-93,KKD-97,DG-04,GM-05,MM-23,
  KB-23,ADG-24,Dober-25}.
In the following, we focus on the most peculiar transition line of the
QAT chain where the critical exponent $\nu$ changes, thus for
$g\approx 1$ and $w\in [-1/\sqrt{2}, 1]$, which contains the weak
intercoupling regime. We will not discuss the other more standard
transition lines of the phase diagram for $w<-1/\sqrt{2}$ and $w>1$.

\subsection{FSS of the correlation functions within ${\cal S}$}
\label{fssgcritline}

To study the critical behavior within ${\cal S}$, we focus on the
correlation functions of the longitudinal $\hat{\sigma}_x^{(1)}$ and
transverse $\hat{\sigma}_x^{(3)}$ spin operators, defined in
Eqs.~(\ref{gxy}) and~(\ref{fxy}), respectively. Note that they are
equal to the analogous correlation functions of the operator
$\hat\tau_x^{(1)}$ and $\hat\tau_x^{(3)}$, because of the
${\mathbb Z}_2$ interchange symmetry
$\hat{\bm \sigma}_{x} \leftrightarrow \hat{\bm \tau}_{x}$
of the QAT model.

\subsubsection{The longitudinal correlation function}
\label{magncorr}

Since $w$ is a marginal parameter along the continuous transition
line, due to the fact that its RG dimension vanishes, the two-point
function along the critical line, i.e., varying $w$ within
$[-1/\sqrt{2},1]$ and keeping $g=1$ fixed, is expected to show the FSS
behavior
\begin{equation}
  G(x_1,x_2;w) \approx L^{-2 y_\phi} {\cal G}(X_1,X_2;w),
  \label{gsca}
\end{equation}
where $X_i \approx x_i/L$ and $y_\phi=\eta/2=1/8$ is the $w$-independent
RG dimension of the order-parameter field corresponding to
$\hat \sigma_{x}^{(1)}$ and $\hat \tau_x^{(1)}$.  In the case of boundary
conditions preserving translational invariance, such as PBC,
$G(x_1,x_2)\equiv G(x_2-x_1)$.  We recall that the scaling function
${\cal G}$ is expected to depend on the boundary conditions.  An
interesting question concerns its dependence on $w$ (i.e., whether,
and how, it depends on $w$).

Some exact results for the scaling behavior of the critical two-point
function can be obtained by CFT approaches (see, e.g.,
Refs.~\cite{It-Dr-book,CFT-book}).  Interestingly, in the case of PBC
preserving translational invariance, no differences are expected in
the FSS of the two-point function $G(x-y) = {\rm Tr} \big[ \hat
  \rho_{\cal S} \, \hat \sigma_x^{(1)} \hat \sigma_y^{(1)} \big]$
along the continuous transition line. Indeed, conformal invariance
strongly constrains its behavior when considering PBC, leading to the
simple behavior
\begin{equation}
  G_{\rm pbc}(x) \propto \left[ L^2 \sin^2(\pi X) \right]^{-y_\phi}.
  \label{gcftpbc}
\end{equation}
Therefore, its critical FSS function
\begin{equation}
  {\cal G}_{\rm pbc}(X) = \left[\sin(\pi X)\right]^{-2y_\phi}
  \label{gscacftpbc}
\end{equation}
depends only on the RG dimension $y_\phi=1/8$ of the order parameter field
(the overall constant is related to an irrelevant
normalization)---see, e.g., Refs.~\cite{It-Dr-book,CFT-book}.  Since
the RG dimension $y_\phi$ remains the same along the continuous
transition line, the FSS of the two-point function for PBC, and its
related quantities turns out to be independent of $w$, i.e., it does
not change along the continuous transition line for $-1/\sqrt{2}<w\le 1$.

This behavior of the two-point function with PBC may suggest that the
FSS along the continuous transition line remains unchanged.  In this
respect we mention that the numerical analyses of Ref.~\cite{MM-23}
did not observe variations of the FSS of the 2D CAT model with PBC
along the transition line where the exponent $\nu$ varies
continuously, putting forward a {\em superuniversality}
hypothesis~\cite{MM-23,Khan-etal-17}, that is that the scaling
functions remain unchanged along the fixed-point line of the CAT model
(up to trivial normalizations).  In the following we show that this
simple scenario is not confirmed by our FSS analysis of the QAT model
with OBC. We recall that, using the quantum-to-classical mapping, the
FSS of the 1D QAT model is expected to be analogous to that of the 2D
CAT model within an infinite slab, with the same boundary conditions
along the finite-size direction~\cite{Sachdev-book,RV-21}.

For OBC, which does not preserve translational invariance, it is more
convenient to rescale the coordinate as $X=x/(L-1)$, especially when
using odd sizes $L=2\ell +1$, and consider the correlation functions
\begin{equation}
  G_0(x) \equiv  G(x_0,x) \approx L^{-2y_\phi} {\cal G}_0(X),
  \quad X={x\over 2\ell},
  \label{g0xdef}
\end{equation}
where $x_0=0$ is the central site of the chain, and $X\in [-1/2,1/2]$.
The FSS of the two-point correlation function $G_0(x)$ with OBC
turns out to be more complex than that with PBC, due to the presence
of the boundaries.  Therefore a numerical analysis of $G_0(x)$
for OBC should provide a nontrivial check of the
$w$-dependence of the critical behaviors along the QAT transition
line. Its exact behavior is known for the critical Ising chain,
showing a more complex structure (see App.~\ref{isichain}).

\begin{figure}[!t]
  \includegraphics[width=0.47\textwidth]{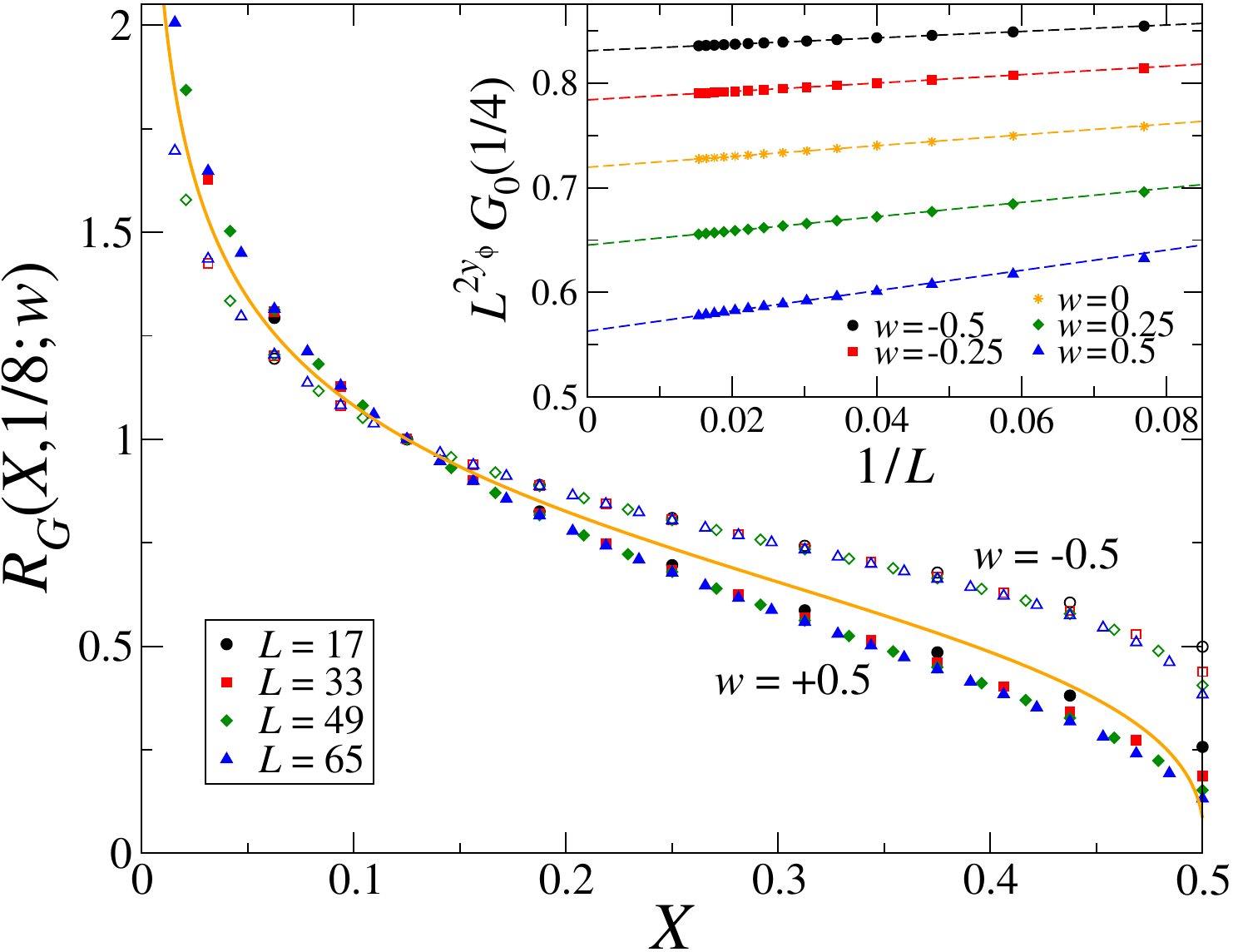}
  \caption{The scaling behavior of the ratio $R_G(x_1=2\ell X,x_2=\ell/4)$,
    defined in Eq.~(\ref{rgdef}). We report data for $w=-0.5$ (empty
    symbols) and $0.5$ (full symbols), up to $L=65$. They confirm
    the approach to asymptotic $w$-dependent FSS curves, as predicted
    by Eq.~(\ref{rgcrit}).  The full line shows the exact result for
    the critical quantum Ising chain ($w=0$), obtained by the CFT
    approach (see App.~\ref{isichain}).  The inset shows results for
    $G_0(\ell/4)$ vs $1/L$, for various values of $w$, confirming
    its asymptotic $L^{-2y_\phi}$ power-law scaling with $1/L$ corrections
    (dashed lines are linear fits to the numerical data
    for $L \geq 25$).}
\label{twopATsca}
\end{figure}

In Fig.~\ref{twopATsca} we show results for the two-point longitudinal
correlaton function, for various values of $w$, as obtained by DMRG
computations up to $L=65$, and compare them with the exact CFT
expressions for the quantum Ising chain ($w=0$), see
App.~\ref{isichain}. For an optimal comparison of the scaling
behaviors for different values of $w$, we plot the ratios
$R_G(X_1,X_2=1/8)$ defined in Eq.~(\ref{rgdef}), which allows us to
eliminate the nonuniversal moltiplicative normalization of the
critical two-point function.  For these quantities, the asymptotic FSS
behavior is expected to be given
\begin{equation}
  R_G(X_1,X_2;w) = R_G^*(X_1,X_2;w) + a_1 L^{-1}
  + a_\omega L^{-\omega} + \ldots
  \label{rgcrit}
\end{equation}
where the $O(L^{-1})$ term is related to the boundary corrections
associated with the OBC, and the $O(L^{-\omega})$ scaling correction
is related to the leading irrelevant RG perturbation at the fixed
point, which may depend on $w$.  We recall that $\omega=2$ for the
Ising chain~\cite{PV-02,CHPV-02}.  The results for the ratios
$R_G(X_1,X_2=1/8;w)$ reported in Fig.~\ref{twopATsca} clearly show
that they approach an asymptotic FSS, as predicted by
Eq.~(\ref{rgcrit}), and their large-$L$ behavior appears to be well
described by $O(L^{-1})$ corrections (see the inset of
Fig.~\ref{twopATsca}), suggesting that $\omega\gtrsim 1$ for all values
of $w$ considered. Moreover, they definitely show that the FSS of the
two-point function with OBC depends on $w$, at variance with the case
of PBC where the FSS (\ref{gscacftpbc}) of the two-point function is
independent of $w$.

\subsubsection{The transverse correlation function}
\label{trasvcorr}

One may also study the scaling behavior of the two-point function of
the transverse spin operator $\hat\sigma_x^{(3)}$, defined in
Eq.~(\ref{fxy}). We expect that its FSS behavior along the QAT
transition line $g=1$ and $w\in [-1/\sqrt{2},1]$ can be written as
\begin{equation}
  F(x_1,x_2;w) \approx  L^{-\kappa} {\cal F}(X_1,X_2;w),
  \label{fxyscaqat}
\end{equation}
with an appropriate exponent $\kappa$ that should be related to some
of the RG operators at the QAT fixed points.  We recall that, at a
standard Ising transition (i.e., when the stacked subsystems are
decoupled), the scaling behavior is the one reported in
Eq.~\eqref{fxysca}, with $\kappa=2\,y_e$ where $y_e = d+z-y_r=1$, due
to the fact that the transverse operator $\hat \sigma_x^{(3)}$ is related
to the energy-density operator at the 2D Ising fixed point.

\begin{figure}[!t]
  \includegraphics[width=0.47\textwidth]{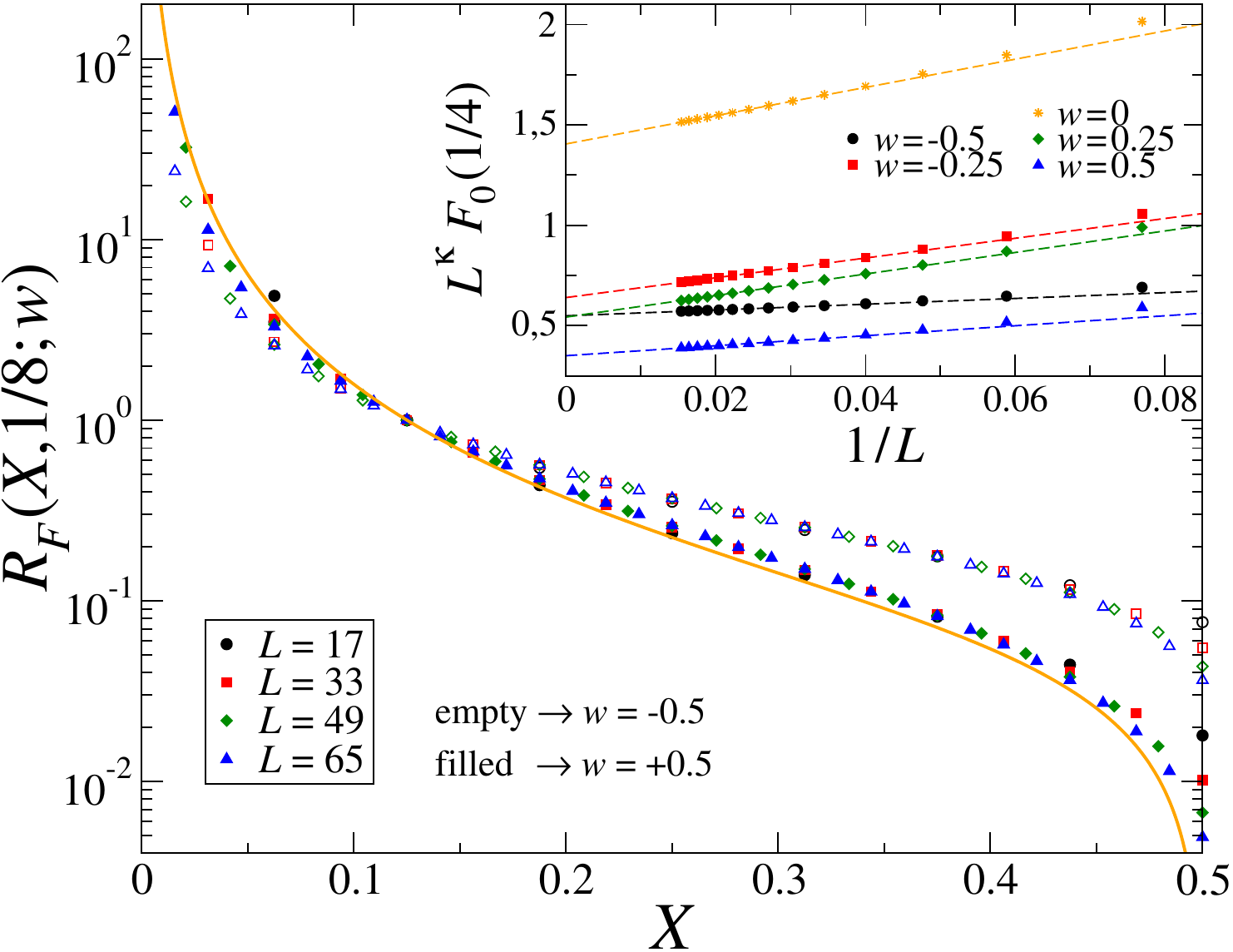}
  \caption{The scaling behavior of the ratio $R_F(X,1/8)$, defined in
    Eq.~\eqref{rfdef}, for $w=-0.5$ and $0.5$.  They confirm the
    approach to asymptotic $w$-dependent FSS curves, as predicted by
    Eq.~(\ref{fxyscaqat}).  The full line shows the exact result
    for the critical quantum Ising chain ($w=0$)---see App.~\ref{isichain}.
    The inset shows results for $F_0(\ell/4)$ vs $1/L$, confirming
    its asymptotic $L^{-\kappa}$ power-law scaling with $1/L$ corrections,
    with $\kappa$ given by Eq.~\eqref{kappahyp} (dashed lines are linear
    fits to the numerical data for $L \geq 25$).}
  \label{Fscaling}
\end{figure}

To check the scaling of the $\hat \sigma_x^{(3)}$ correlations, we also
define the correlations $F_0(x)=F(x_0,x)$ with the central point of
the lattice, and the ratios
\begin{equation}
  R_F(X_1,X_2) = {F_0(x_1=2 \ell X_1)\over F_0(x_2=2\ell X_2)}.
  \label{rfdef}
\end{equation}
Our numerical analysis confirms the scaling behavior reported in
Eq.~(\ref{fxyscaqat}), as demonstrated by the data for the ratio
$R_F(X,1/8)$ shown in Fig.~\ref{Fscaling}, which appear to approach
asymptotic $w$-dependent scaling curves.  However, this check does not
provide information on the value of $\kappa$ itself. The determination
of the exponent $\kappa$ in the QAT correlation function $F(x,y)$
turns out to be more complicated than that at the quantum Ising
transitions, because there are two independent RG operators
associated with $\hat \sigma_x^{(3)}$.  Indeed, we may write
\begin{align}
  & \hat\sigma_x^{(3)} = \hat E_x + \hat C_x,  \label{sigdec}\\
  & \hat E_x = \hat\sigma_x^{(3)} + \hat\tau_x^{(3)},\quad
  \hat C_x =   \hat\sigma_x^{(3)} - \hat\tau_x^{(3)},
\label{ecdef} 
\end{align}
where $\hat E_x$ and $\hat C_x$ are operators associated with
different RG perturbations~\cite{Kadanoff-77,KNK-81,DK-82} of the
$w$-dependent fixed points along the QAT transition line. The operator
$\hat E_x$ can be associated with the energy RG perturbation
controlling the critical behavior when changing $g$, thus
\begin{equation}
  y_e = d + z - \frac{1}{\nu} = {2\nu-1\over \nu},
  \label{yedef}
\end{equation}  
with the $w$-dependent exponent $\nu$ reported in Eq.~(\ref{nuexp}).
The other operator $\hat C_x$, called crossover
operator~\cite{KNK-81}, has a different RG dimension given
by~\cite{Kadanoff-77,KNK-81}
\begin{equation}
  y_c= y_e^{-1}= { \nu \over 2\nu - 1}.
  \label{ycdef}
\end{equation}
The asymptotic FSS behavior of the $\hat\sigma_x^{(3)}$ correlation
function (\ref{fxyscaqat}) must be controlled by the smallest RG
dimension between $y_e$ and $y_c$, therefore the exponent $\kappa$ in
Eq.~(\ref{fxyscaqat}) should be given by~\footnote{This can be
understood by recalling that the critical spatial correlations of a
generic operator $\hat{O}(x)$ decay as $G_O(x,y) \sim |x-y|^{-2
  y_o}$. Therefore the correlations of operators with smaller RG
dimensions $y_o$ are less suppressed in the large-distance limit.}
\begin{equation}
  \kappa = 2 \, {\rm Min}[y_e,y_c].
  \label{kappahyp}
\end{equation}
Notice that $y_e=y_c$ when $\nu=1$, which corresponds to the case of
the Ising universality class, thus for $w=0$. Therefore, we have
$\kappa = 2 y_e$ for $w>0$ and $\kappa = 2 y_c$ for $w<0$.  For
example, this would imply that $\kappa=3/2$ for $w=0.5$ and
$\kappa=4/3$ for $w=-0.5$. The formula for $\kappa$ reported in
Eq.~(\ref{kappahyp}) is nicely confirmed by our numerical
simulations, see, for example, the inset of Fig.~\ref{Fscaling}, which
shows data for the correlation $F(0,\ell/4)$ for various values of
$w$.

\subsubsection{Other correlation functions}
\label{fsscritline}

We have also computed the FSS behavior of correlation functions
involving the spin operator of the environment ${\cal E}$.
In partcular, we verified the FSS of the connected correlation
function of the energy operator $\hat E_x$ defined in Eq.~(\ref{ecdef}),
\begin{equation}
  E(x_1,x_2) \!=\! \langle\Psi_0| \hat E_{x_1}\hat E_{x_2} |\Psi_0 \rangle -
  \langle\Psi_0| \hat E_{x_1}|\Psi_0 \rangle \, \langle\Psi_0| \hat E_{x_2} |\Psi_0 \rangle,
  \label{exy}
\end{equation}
where $|\Psi_0\rangle$ is the ground state of the QAT model.
Numerical data (not shown) nicely support the expected asymptotic FSS
behavior
\begin{equation}
  E(x_1,x_2;w) \approx  L^{-2y_e} {\cal E}(X_1,X_2;w),
  \label{exyscaqat}
\end{equation}
with $y_e$ given in Eq.~(\ref{yedef}). Note that this scaling behavior
holds even for negative values of $w$, unlike the case of the
transverse correlation function of the operator $\hat\sigma_x^{(3)}$
[see Eq.~(\ref{fxyscaqat}) with $\kappa$ given in Eq.~(\ref{kappahyp})].

\begin{figure}[!t]
  \includegraphics[width=0.47\textwidth]{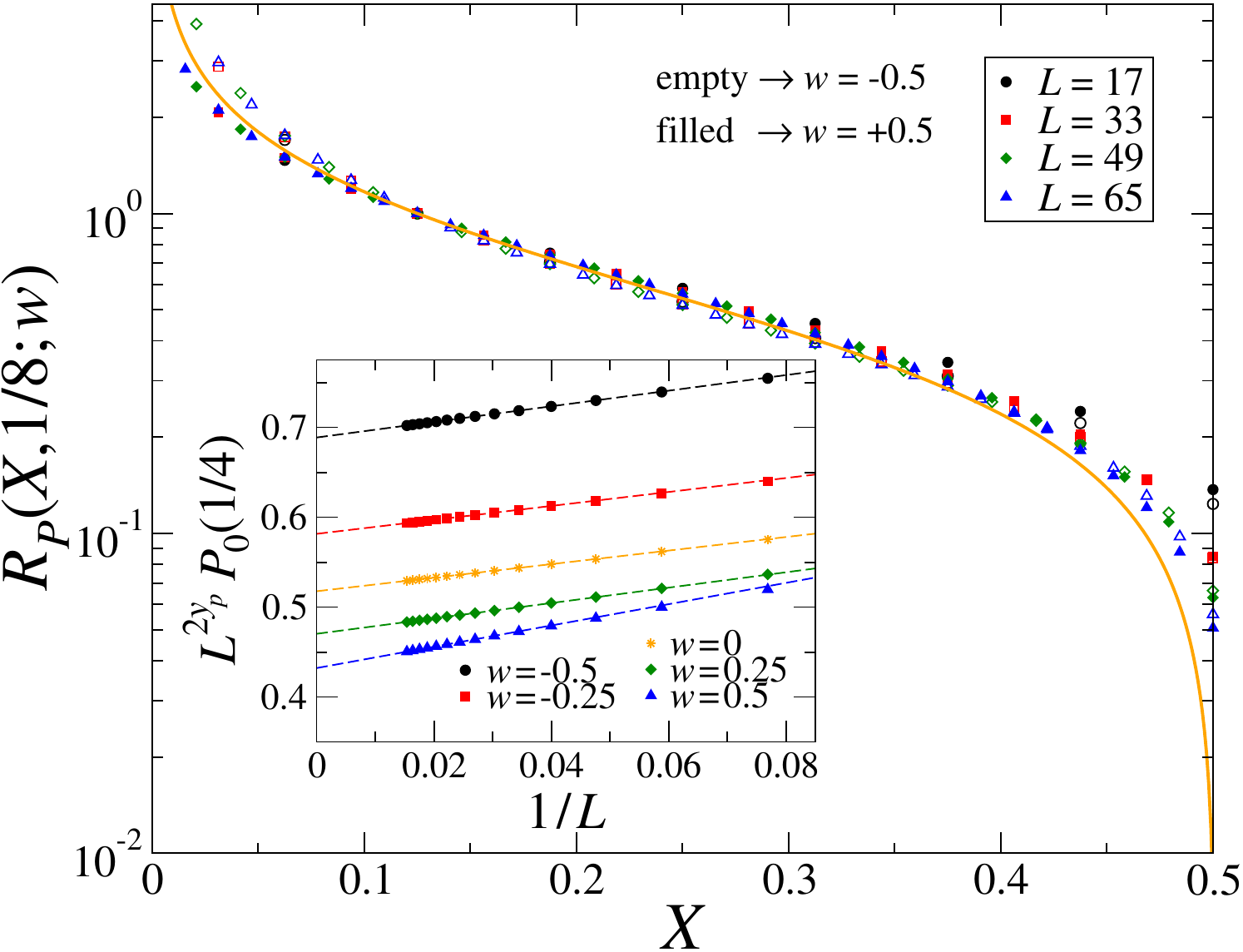}
  \caption{The scaling behavior of the ratio $R_P(X,1/8)$, defined in
    Eq.~(\ref{rwdef}), for $w=-0.5$ and $0.5$.  They confirm the
    approach to asymptotic $w$-dependent FSS curves, as predicted by
    Eq.~(\ref{wsca}).  The full line shows the exact result for the
    critical quantum Ising chain ($w=0$)---see App.~\ref{isichain}.
    The inset shows results for $P_0(\ell/4)$ vs $1/L$, confirming
    its asymptotic $L^{-2y_p}$ power-law scaling with $1/L$ corrections
    (dashed lines are linear fits to the numerical data for $L \geq 25$).}
\label{Wscaling}
\end{figure}

We finally consider the correlation function of the polarization
operator $\hat P_x=\hat{\sigma}_x^{(1)}\hat{\tau}_{x}^{(1)}$, i.e.,
\begin{equation}
  P(x_1,x_2) = \langle\Psi_0| \hat P_{x_1} \hat P_{x_2} |\Psi_0 \rangle.
  \label{axy}
\end{equation}
Its FSS behavior is expected to be
\begin{eqnarray}
  &&P(x_1,x_2;w) = L^{-2 y_p} {\cal P}(X_1,X_2;w),  \label{wsca}\\
  &&y_p= {y_e\over 4} = {1\over 2} - {1\over 4\nu}, 
  \label{ypdef}
\end{eqnarray}
where $y_p$ is the RG dimension of the operator
$\hat P_x$~\cite{WL-74,DR-79,KNK-81,DK-82,MM-23}, which depends on $w$ as
well (for example, $y_p= 3/8$ for $w=-1/2$ and $y_p=3/16$ for
$w=1/2$).  Note that, for $w=0$ (i.e., when the stacked Ising subsystems get
decoupled), the correlation function $P$ equals the square of the
Ising-chain two-point function of the longitudinal spin operator
$\hat \sigma_x^{(1)}$. Therefore, we must have $y_p = 2 y_\phi= 1/4$ for
$w=0$, which is consistent with the value obtained by inserting the
Ising value $\nu=1$ into Eq.~(\ref{ypdef}).  Analogously to the other
correlation functions, we define the correlations $P_0(x)=P(x_0,x)$
with the central point of the lattice, and the ratios
\begin{equation}
  R_P(X_1,X_2) = {P_0(x_1=2 \ell X_1)\over P_0(x_2=2\ell X_2)}.
  \label{rwdef}
\end{equation}
As shown in Fig.~\ref{Wscaling}, the FSS behavior (\ref{wsca}) of the
$P$-correlation is nicely confirmed by the data.

\subsection{FSS of the ground-state fidelity}
\label{gssgfid}

We finally discuss the FSS behavior of the ground-state fidelity when
changing the Hamiltonian parameters along the fixed-point line of the
QAT model. In fact, the quantum fidelity is able to provide
information of the quantum critical correlations developed by the
system~\cite{AFOV-08, Gu-10, BAB-17, RV-21}.  It is a geometrical
object that can be used to monitor the changes in the ground-state
wave function $|\Psi_0(u,L)\rangle$ when one of the control
parameters, say $u$, is varied by a small amount $\varepsilon$.  The
ground-state fidelity associated with a generic Hamiltonian parameter
$u$ is defined as
\begin{equation}
  A(u,\varepsilon,L) \equiv | \langle \Psi_0(u+\varepsilon,L) | \Psi_0(u,L)
  \rangle|.
  \label{fiddef}
\end{equation}
Of course, $A(u,0,L) = 1$.
Assuming $\varepsilon$ to be sufficiently small, one can expand
Eq.~\eqref{fiddef} in powers of $\varepsilon$~\cite{Gu-10}:
\begin{equation}
  A(u,\varepsilon,L) 
  = 1 - {1\over 2}\,\varepsilon^2 \, \chi_A(u,L) + O(\varepsilon^3),
  \label{expfide}
\end{equation}
where $\chi_A$ defines the fidelity susceptibility (the cancellation
of the linear term of the expansion is essentially related to the fact
that the fidelity is bounded, i.e., $F\le 1$).  Under normal conditions,
the fidelity susceptibility is expected to grow proportionally to the
volume, i.e., $\chi_A\sim L^d$.  However, its power law can change
drastically across a continuous transition driven by the Hamiltonian
parameter $u$ with RG dimension $y_u$.
Indeed, one expects the behavior~\cite{RV-21,RV-18}
\begin{equation}
  \chi_A(u,L) \approx L^{2 y_u} {\cal A}_2(u L^{y_u}).
    \label{chiasca}
\end{equation}
Note, however, that Eq.~\eqref{chiasca} describes the leading scaling
behavior only when $2y_u>d$. Otherwise, for $2y_u<d$, the power-law with
increasing size is dominated by the analytic contributions at
the continuous transition~\cite{RV-21}, i.e., $\chi_A\sim L^d$, while
the scaling term in Eq.~\eqref{chiasca} provides only a subleading
contribution.

We may define an analogous ground-state fidelity susceptibility
$\chi_A(w,L)$ along the critical line of QAT model (thus for $g=1$),
with respect to variations of the interaction parameter $w$.
Since $w$ is marginal along the critical line, meaning that its RG
dimension $y_w$ vanishes, we would expect the normal asymptotic
large-size behavior
\begin{equation}
  \chi_A(w,L) \sim L.
  \label{chiascaw}
\end{equation}
On the other hand, keeping $w$ fixed and varying $g$ around its
critical value $g=g_c=1$, whose RG dimension is $y_g=1/\nu$, we expect
the asymptotic large-size behavior
\begin{equation}
  \chi_A(g,L) \approx L^{2/\nu} {\cal A}_2 \big[ (g-g_c) L^{1/\nu} \big]
  \quad {\rm for}\;\;\nu < 2, \label{chiascag}
\end{equation}     
therefore for $-1/2< w \le 1$. On the other hand,
$\chi_A(g,L) \sim L$ for $\nu \ge 2$,
corresponding to $-1/\sqrt{2} < w \le -1/2$.

\section{Two-dimensional stacked quantum Ising systems}
\label{d2stacked}

In this section, we extend the discussion to higher-dimensional SQI
models, in particular focusing on 2D SQI systems, while still assuming local
interactions that preserve the ${\mathbb Z}_2$ symmetry of each quantum
Ising subsystem.  These systems are defined analogously to the SQI chains,
by extending the Hamiltonian terms~\eqref{twoisiham1} to higher dimensions,
i.e.,
\begin{subequations}
  \label{twoisihamgd}
  \begin{align}
    & \hat H_\sigma = - J \sum_{\langle {\bm x}
      {\bm y}\rangle} \hat\sigma_{\bm x}^{(1)} \hat\sigma_{\bm y}^{(1)}
    - g \sum_{\bm x} \hat\sigma_{\bm x}^{(3)} , \\
    & \hat H_\tau = - J_e \sum_{\langle {\bm x} {\bm y}\rangle}
    \hat\tau_{\bm x}^{(1)} \hat\tau_{\bm y}^{(1)}
    - g_e \sum_{\bm x} \hat\tau_{\bm x}^{(3)} , \\
    & \hat H_w = - w \Bigg[ \sum_{\langle {\bm x} {\bm y}\rangle}
      \hat\sigma_{\bm x}^{(1)} \hat\sigma_{\bm y}^{(1)}
      \hat\tau_{\bm x}^{(1)} \hat\tau_{\bm y}^{(1)} +
    \sum_{\bm x} \hat\sigma_{\bm x}^{(3)} \hat\tau_{\bm x}^{(3)}\Bigg],
  \end{align}
\end{subequations}
where $\hat\sigma_{\bm x}^{(k)}$ and $\hat\tau_{\bm x}^{(k)}$ are two
sets of Pauli matrices ($k=1,2,3$) on the site ${\bm x}$,
${\langle {\bm x} {\bm y}\rangle}$ denotes nearest-neighbor sites.
As before, we consider $J=J_e=1$.

Again, for $w=0$, we have two decoupled quantum Ising systems. We
recall that quantum Ising models in $d\ge 2$ dimensions undergo a
continuous transition as well, at a finite value $g_c$, belonging to
the $(d+1)$-dimensional Ising universality class.  Accurate estimates of
the critical exponents of the 3D Ising universality class have been
obtained using various approaches (see, e.g.,
Refs.~\cite{PV-02,GZ-98,CPRV-02,Hasenbusch-10,KPSV-16,KP-17,
  FXL-18,Hasenbusch-21}); in particular~\cite{KPSV-16},
$\nu=0.629971(4)$.  The critical exponents of the 4D Ising
universality class take mean-field values, $\nu=1/2$ and $\eta=0$,
with additional multiplicative logarithmic corrections~\cite{PV-02}.
The dynamic exponent $z$ is 1 in any dimension.

The scenario in which the subsystem ${\cal S}$ is close to criticality
and weakly coupled to an environment ${\cal E}$ far from
criticality is analogous to that discussed in Sec.~\ref{smallw}, with
transition lines belonging to the $(d+1)$-dimensional Ising
universality class. Therefore, we focus on the case where both SQI
subsystems are critical, leading to a distinct multicritical scenario
in which an effective enlargement of the symmetry
from ${\mathbb Z}_2\oplus{\mathbb Z}_2$ to O(2) takes place,
thus differing substantially from that found for SQI chains.

\begin{figure}[!t]
  \includegraphics[width=0.8\columnwidth]{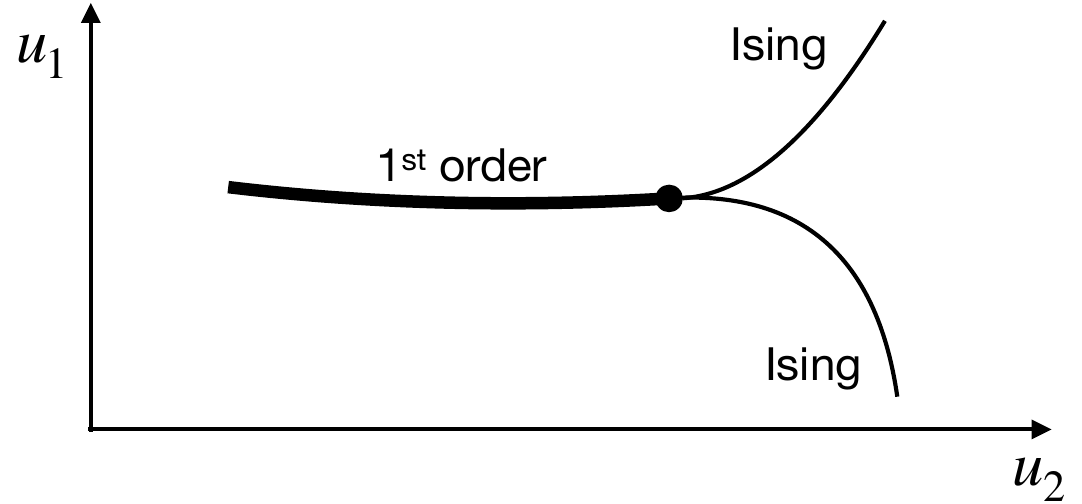}
  \caption{Sketch of a phase diagram containing a 3D bicritical XY
    point arising from the competition of two distinct Ising-like
    order parameters, for generic system parameters $u_1$ and $u_2$.
    The multicritical behavior at the bicritical XY point is described
    by the 3D LGW theory (\ref{bicrHH}).  It is located at the meeting
    point of three transitions lines, one first-order transition line
    (thick full line) and two continuous Ising transition lines (full
    lines) whose order parameters are related to the real scalar
    fields $\varphi_1$ and $\varphi_2$ of the LGW theory
    (\ref{bicrHH}). At the bicritical XY point the symmetry of the
    critical modes gets effectively enlarged, from ${\mathbb Z}_2
    \oplus {\mathbb Z}_2$ to O(2).}
  \label{sketchbicr}
\end{figure}

In these conditions, the systems exhibit multicritical behaviors
characterized by two relevant RG perturbations, both invariant under
the global symmetry of the model and both of which must be
tuned to reach the multicritical point. We indicate with $r_1$
and $r_2$ the corresponding parameters, normalized such that $r_1=r_2=0$
at the multicritical point. Of course, they must be functions of the
SQI Hamiltonian parameters $g$, $g_e$, and $w$.  Assuming, reasonably,
that the dynamic exponent at the multicritical point is $z=1$,
the competition between the order parameters of $d$-dimensional SQI
models can be described by an effective $(d+1)$-dimensional
Landau-Ginzburg-Wilson (LGW) $\Phi^4$ theory. This theory
involves two real order-parameter fields, $\varphi_1$ and $\varphi_2$,
which are invariant under a ${\mathbb Z}_2\oplus{\mathbb Z}_2$ symmetry,
i.e., under the sign change of each field, so that only even powers of
each field are allowed.  The corresponding Hamiltonian
is~\cite{LF-72,FN-74,NKF-74}
\begin{eqnarray}
  {\cal H} & = & (\partial_\mu \varphi_1)^2 + (\partial_\mu \varphi_2)^2 +
  r_1 \, \varphi_1^2 + r_2 \, \varphi_2^2 \nonumber \\
  & & + \,v_1\, \varphi_1^4 + v_2 \,\varphi_2^4 + v_3 \,\varphi_1^2\varphi_2^2.
  \label{bicrHH} 
\end{eqnarray}
Mean-field analyses show that this theory admits a bicritical
point~\cite{LF-72,FN-74,NKF-74}, as sketched in Fig.~\ref{sketchbicr}.
The 3D RG flow of the quartic
couplings determines the 3D multicritical behavior when tuning the
quadratic parameters $r_1$ and $r_2$ to the multicritical point.  The
stable fixed point of the ${\mathbb Z}_2\oplus{\mathbb Z}_2$
multicritical LGW theory (\ref{bicrHH}) turns out to be the
O(2)-symmetric XY fixed point~\cite{NKF-74, CPV-03, HV-11, BPV-22-z2g}.
Therefore, the multicritical behavior at the bicritical point of 2D
${\mathbb Z}_2\oplus{\mathbb Z}_2$ symmetric quantum systems must
generally belong to the XY universality class,\footnote{Accurate
estimates of the critical exponents characterizing the 3D XY
universality class can be found in
Refs.~\cite{PV-02,CHPV-06,HV-11,CLLPSSV-20,BPV-25,Hasenbusch-25,
  Hasenbusch-25-2}, obtained by various theoretical approaches.} thus
realizing an effective enlargement of the symmetry of the
multicritical modes, from ${\mathbb Z}_2\oplus{\mathbb Z}_2$ to O(2).

For values of the QAT parameter $w$ that are within the attraction
domain of the bicritical XY point $g=g_e=g_{mc}$ (where $g_{mc}$
generally depends on $w$), the singular part of the free-energy
density is expected to develop the infinite-volume scaling
behavior~\cite{FN-74,NKF-74,PV-02,CPV-03,BPV-22-z2g}
\begin{equation}
  F_{\rm sing}(u_1,u_2) = |u_2|^{d/y_2} {\cal F}_\pm (u_1 |u_2|^{-\phi}),
  \label{freeenmcp2}
\end{equation} 
in the infinite-volume limit and neglecting subleading corrections,
where $u_1$ and $u_2$ are the nonlinear scaling fields associated with
the two relevant parameters $r_1$ and $r_2$, and $y_1>0$ and $y_2>0$
are the corresponding RG dimensions at the multicritical XY fixed
point. The scaling fields $u_1$ and $u_2$ are analytic functions of
LGW $\Phi^4$ theory (\ref{bicrHH}), and therefore of the SQI
parameters $g$, $g_e$, and $w$.  In the above scaling equation, we
neglected corrections to the multicritical behavior due to the
irrelevant scaling fields.  The functions ${\cal F}_\pm(X)$ apply to
the parameter regions in which $\pm u_2 > 0$, respectively.  The
bicritical point is the starting point of three transition lines: two
Ising transition lines and one first-order transition line. They
follow the scaling equation $X = u_1 |u_2|^{-\phi} = {\rm const}$ with
a different constant for each transition line.  Since $\phi > 1$, they
are tangent to the line $u_1 = 0$ (see Fig.~\ref{sketchbicr}).

Several field-theoretical and numerical works have determined the
exponents $y_i$ entering the scaling ansatz associated with a
bicritical XY point. The leading RG exponents $y_1$ and $y_2$
correspond to the RG dimensions of the quadratic spin-2 and spin-0
perturbations at the XY fixed point. The leading RG exponent $y_1$ is
associated with the quadratic spin-two perturbation, whose RG
dimension is~\cite{Hasenbusch-25-2} $y_1 = 1.76370(12)$.  The second
largest exponent is associated with the spin-zero quadratic operator
and is directly related to the correlation-length critical exponent at
standard XY transitions: $y_2 =1/\nu_{xy}=1.48872(5)$, obtained using
the estimate~\cite{Hasenbusch-25} $\nu_{xy}=0.671718(23)$.  Using the
above results, we can estimate the crossover exponent
$\phi = y_1/y_2 = 1.1847(1)$.  Other details on the scaling correction
exponents, which are related to the spin-4, spin-2, and spin-zero
quartic perturbations, respectively, can be found
in Refs.~\cite{PV-02,CPV-00,CPV-03,HV-11,BPV-22-z2g,Hasenbusch-25-2}.

Concerning the 2D SQI model, for values of the intercoupling parameter
$w$ within the attraction domain of the multicritical XY fixed point,
the multicritical point is located at $g=g_e=g_{\rm mc}$ (where
$g_{\rm mc}$ is expected to depend on $w$) and the scaling fields
entering the scaling Eq.~(\ref{freeenmcp2}) can be identified as
$u_1 \approx g - g_e$ (the first-order transition line departing from
the bicritical XY point is expected to run along the line $u_1=0$) and
$u_2$ along any other direction (in particular, one may choose
$u_2 \approx g +g_e-2 g_{\rm mc}$).  Then, when approaching the
multicritical point along the line $u_1=0$, the correlation length is
expected to increase as $\xi \sim u_2^{-\nu_{xy}}$, while along any
other direction it increases as $\xi\sim u_1^{-1/y_1}$.  Moreover, at
the multicritical point, the algebraic decay of the two-point function
$G({\bm x},{\bm y})$ of the longitudinal operator
$\hat \sigma_{\bm x}^{(1)}$ [defined analogously to that of the 1D model,
  cf.~Eq.~(\ref{gxy})] is controlled by the 3D XY exponent
$\eta_{xy} = 0.03816(2)$~\cite{Hasenbusch-25}, i.e.,
$G({\bm x},{\bm y})\sim |{\bm x}-{\bm y}|^{-1-\eta_{xy}}$.

\section{Conclusions}
\label{conclu}

We have addressed the quantum behavior of a many-body system ${\cal S}$
interacting with a surrounding many-body environment ${\cal E}$,
assuming that the global system is in its ground state.  As
paradigmatic models, we consider SQI systems, and in particular SQI
chains described by Hamiltonian~\eqref{twoisiham} (see
Fig.~\ref{sketchsystem}). One chain plays the role of the open system
${\cal S}$ under observation, while the other one acts as the
environment ${\cal E}$. The two subsystems interact by means of the
Hamiltonian term $\hat H_w$ defined in Eq.~\eqref{Hsigtau}, which
preserves the ${\mathbb Z}_2$ symmetries of the SQI subsystems.

We analyze quantum correlations within subsystem ${\cal S}$
and study their dependence on the state of the weakly-coupled
complementary part ${\cal E}$ and on the intercoupling strength.
We focus on the quantum critical regimes of ${\cal S}$, showing
that the corresponding scaling behaviors crucially depend on whether
${\cal E}$ is itself critical or far from criticality.
The most interesting regime arises when both SQI subsystems develop
critical correlations. In particular, we consider the case in which
the two SQI subsystems are identical, so that the global system is
equivalent to the QAT
model~\cite{KNK-81,IS-84,AF-84,Fradkin-84,Shankar-85,GR-87,ABB-88},
being characterized by an additional ${\mathbb Z}_2$ interchange
symmetry between the subsystems. In this limit, 1D SQI systems develop
a peculiar critical line where the length-scale critical exponent
$\nu$ depends on the intercoupling strength $w$.  Interesting
phenomena also occur in higher dimensions.  Indeed, we argue that 2D
SQI systems display multicritical behaviors characterized by an
enlargement of the symmetry of critical modes, from the actual
${\mathbb Z}_2 \oplus {\mathbb Z}_2$ symmetry to a continuous O(2)
symmetry.

The SQI model studied in this paper substantially differs from the one
analyzed in Ref.~\cite{FPV-23}, where the Hamiltonian term is
proportional to the product of the longitudinal spin variables [see
Eq.~\eqref{Hkint}], thus breaking the ${\mathbb Z}_2$ symmetries of
the individual SQI systems and leaving only a global residual
${\mathbb Z}_2$ symmetry. Indeed, substantial differences emerge when
comparing the critical behaviors of SQI models that preserve or break
the original symmetries of the decoupled subsystems.
The results for the SQI systems considered here, together with those
reported in Ref.~\cite{FPV-23}, clearly demonstrate that the quantum
behavior of ${\cal S}$ (including its coherence properties, phase
diagram, and critical behavior) depends strongly on the quantum state
of the environment ${\cal E}$.  However, the specific features of the
resulting critical behaviors may also depend crucially on the nature
of interactions between ${\cal S}$ and ${\cal E}$, and in particular
on which symmetries of the two subsystems are preserved.

Most of the arguments developed here can be straightforwardly extended
to cases in which ${\cal S}$ and ${\cal E}$ are of different nature,
dimensionality, etc, including situations where ${\cal E}$ is much
larger than the observed system ${\cal S}$. Furthermore, the results
obtained for weakly coupled $d$-dimensional SQI systems can be
extended to the corresponding classical systems, i.e., to coupled
$(d+1)$-dimensional classical Ising systems with intercoupling that
preserve the ${\mathbb Z}_2$ symmetries of the two subsystems.

The effects of interactions with a bath (environment) on quantum
many-body systems have also been investigated using different
approaches (see, e.g., Refs.~\cite{Weiss-book, CL-83, LCD-87, WTS-04,
  SWT-04, WVTC-05, YMZ-14, ARBA-17, KMSFR-17, WSBR-18, NRV-19, RV-20,
  RV-21}).  A widely used framework is based on the Lindblad master
equation~\cite{BP-book, RH-book}, which describes certain classes of
dissipative interactions, without keeping track of the full
environment dynamics.  As shown in various studies within the Lindblad
framework, the coupling to the environment generally makes the quantum
critical behavior of a closed system unstable~\cite{YMZ-14, NRV-19,
  RV-20, RV-21}, similarly to finite-temperature effects. In this
context, dissipative couplings act as relevant perturbations that
drive the system away from the quantum critical behavior of the
isolated model.  Alternatives for dissipation are provided by coupling
a many-body system to an infinite set of harmonic oscillators (see,
e.g., Refs.~\cite{Weiss-book, CL-83, LCD-87, WTS-04, SWT-04,
  WVTC-05}), or to a quantum measurement apparatus (see, e.g.,
Refs.~\cite{LCF-18, LCF-19, CTD-19, SRN-19, RV-20B, TPDC-24, MTC-25}).
These types of dissipative interactions are likewise relevant
perturbations of the critical behavior of isolated systems and may
lead to different forms of dissipation-driven criticality, such as
those arising at finite temperature.  We emphasize that our setting is
substantially different: the global system is isolated and evolves
unitarily, rather than being governed by a reduced dissipative
dynamics.  As a result, our scenario is qualitatively distinct, still
allowing for the observation of 1D quantum critical behaviors in
systems with short-range interactions.

\appendix

\section{Details on the DMRG simulations}
\label{app:DMRG}

To determine the ground-state properties of the two SQI chains under
investigation here, we employ a standard two-site DMRG algorithm in
its finite-system formulation~\cite{Schollwock-05}.  We push our DMRG
simulations up to $L=65$, corresponding to a total number of
$2L = 130$ sites. To ensure convergence we perform five finite-system
back-and-forth sweeps for each parameter set.  A crucial parameter in
such kind of simulations is the number $m$ of states (i.e., the
bond-link dimension) retained in the renormalization procedure of the
enlarged block density matrix.  We choose $m=100$ for $L \leq 29$,
$m=120$ for $33 \leq L \leq 49$, and $m=140$ for $53 \leq L \leq 65$,
after carefully checking convergence for all the considered system
sizes.

\begin{figure}[!b]
  \includegraphics[width=0.47\textwidth]{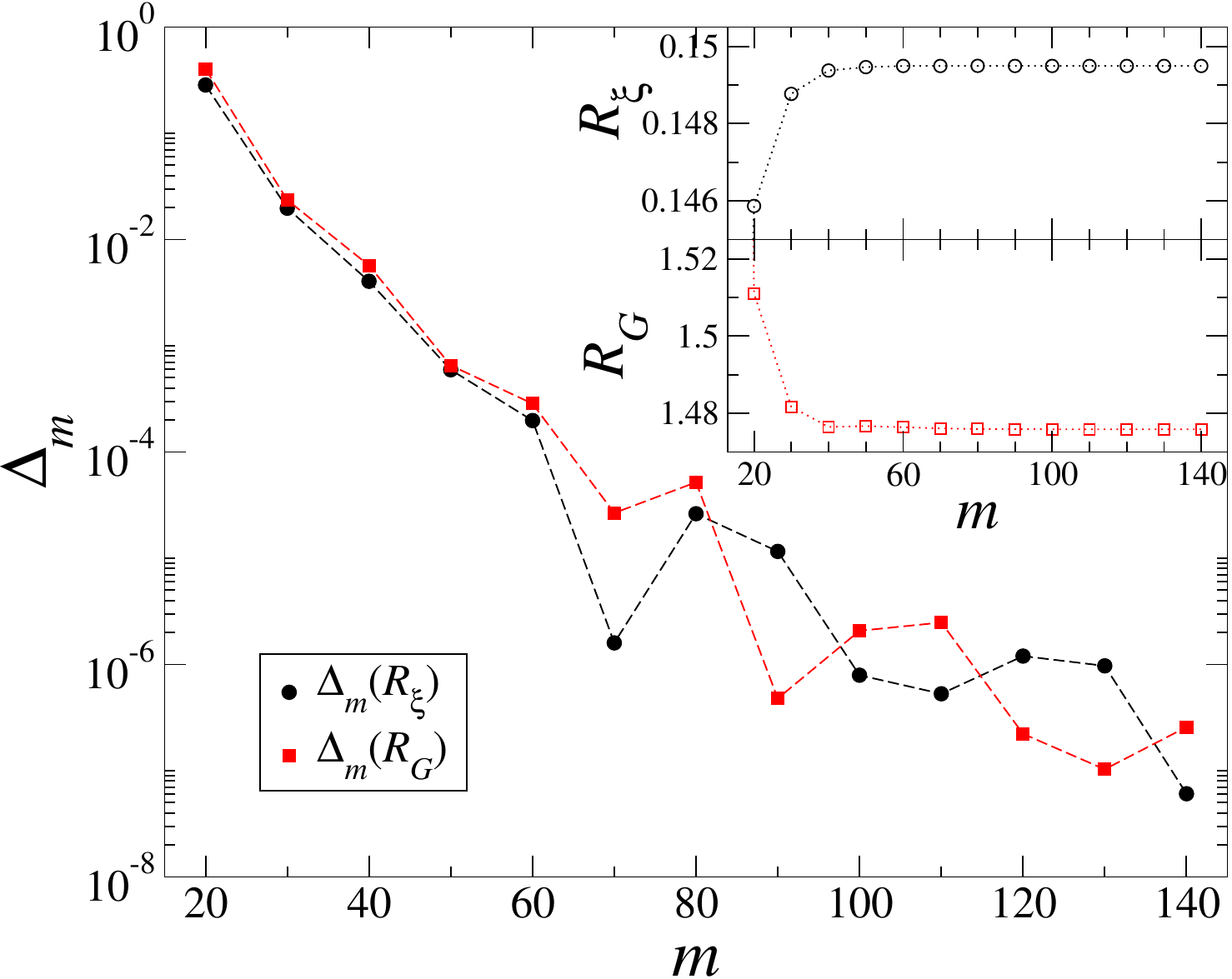}
  \caption{The symmetric relative difference $\Delta_m$ for our DMRG
    calculations of $R_\xi$ (black) and $R_G(1/8,1/4)$ (red), computed
    by increasing the number of kept states in steps of $\Delta m =
    10$.  The two insets show the explicit dependence of $R_\xi$ and
    $R_G$ on $m$.  Data are for the 1D QAT model with $J=J_e=1$,
    $g=g_e=1$, and $w=+0.5$. Here we fix the system size as $L=65$.}
\label{ConvDMRG}
\end{figure}

The dependence of our results on $m$ is illustrated
in Fig.~\ref{ConvDMRG}, where we focus on the largest system size
$L=65$, for parameters chosen such that the system is close to criticality.
To quantify convergence, we consider the symmetric relative
difference
\begin{equation}
  \Delta_m(O) = \frac{2 \vert O(m) - O(\bar m) \vert}
    {O(m) + O(\bar m)}
\end{equation}
for a given monitored observable $O>0$, comparing results obtained by
keeping $m$ states and $\bar m$ states.  In the main frame of the
figure, each data point corresponds to $m$ ranging from $20$ to $140$,
with $\bar m = m - \Delta m$ and $\Delta m = 10$.  For both the
correlation length per site $R_\xi$ [cf.~Eq.~\eqref{rxidef}] and the
ratio $R_G(1/8,1/4)$ of the longitudinal correlation function $G$ at
different distances [cf.~Eq.~\eqref{rgdef}], we observe an exponential
suppression of $\Delta_m$ with increasing $m$.  Choosing $m = 140$
ensures that relative differences remain below $10^{-6}$, for all the
monitored observables.  The two insets clearly show that discrepancies
between the results obtained at different and large values of $m$ are
barely visible on the scale of the figure.  Likewise, we estimate
systematic errors in all the quantities reported in the main text to
be smaller than the symbol size.

\section{Some CFT results for the critical correlations of
  the quantum Ising chain}
\label{isichain}

Here we focus on a single quantum Ising chain at the critical point
with OBC, described by the Hamiltonian~\eqref{Hsig} with $J=g=1$.
We recall that the particular case of the QAT model for $w=0$
corresponds to two decoupled quantum Ising chains. Therefore, in that
case, the ground-state two-point function of the quantum Ising chain
corresponds to the one defined in Eq.~\eqref{gxy}.

The 2D Ising universality class can be associated with a CFT with
central charge $c=1/2$.  CFT provides the asymptotic FSS behavior of
the two-point function at the critical point (see, e.g.,
Refs.~\cite{It-Dr-book,CFT-book}). We report some useful formulas
for the critical two-point function for $L\times\infty$ with OBC,
i.e., with coordinates $-\ell\le x \le \ell$ and
$y \in {\mathbb R}$. Thus $L=2\ell+1\approx 2\ell$ asymptotically in
the large-$L$ limit.

Setting $\vec{r}_i\equiv (x_i,y_i)$, $z_i \equiv x_i+\ell$,
$Z_\pm\equiv (z_2\pm z_1)/L$, $Y \equiv (y_2-y_1)/L$, the critical
two-point function on a infinite strip of transverse size $L$ along
the $z$ axis and OBC reads
\begin{align}
  &G_{\rm obc}(\vec{r}_1,\vec{r}_2) = { \pi^{1/4}\over
    \left[ 4 L^2 \sin\left({\pi z_2/L} \right)
      \sin\left({\pi z_1/L}\right)\right]^{1/8} }
  \label{gcftobcis}\\
  & \times \!\! \left[ {|\sin[\pi(Z_++iY)/2]|^{1/2}\over |\sin[\pi(Z_-+iY)/2]|^{1/2}}
    \!-\! {|\sin[\pi(Z_-+iY)/2]|^{1/2}\over |\sin[\pi(Z_++iY)/2]|^{1/2}} \right]^{1/2}\!.
  \nonumber
\end{align}
The fixed-time ground-state two-point function $G(x,y)$ of the quantum
Ising chain at the critical point is obtained by setting $Y=0$ in
Eq.~(\ref{gcftobcis}).  Note that one should allow for a
multiplicative normalization when comparing $G_{\rm obc}$ with lattice
computations at the critical point in the large-$L$ limit keeping
$Z_{\pm}$ and $Y$ fixed.

This result allows us to exactly compute the universal large-$L$ limit
of some RG invariant quantities, such as $R_\xi$ and $R_G$ defined in
Eqs.~\eqref{rxidef} and~\eqref{rgdef}, respectively.  Using
Eq.~\eqref{gcftobcis}, we obtain the critical values~\cite{CV-10,CPV-14}
\begin{equation}
  R_\xi^* \approx 0.159622, \qquad R_G^*(1/8,1/4) \approx 1.358609.
  \label{rstarobc}
\end{equation}

For the Ising chain, we may also consider the connected equal-time
two-point function of the operator $\hat \sigma^{(3)}_x$ [cf.~Eq.~\eqref{fxy}].
CFT computations give~\cite{It-Dr-book,CV-10}
\begin{equation}
  F_{\rm obc}(0,x) \sim { {\rm cos}\left({\pi x/L}\right) \over L^2 \,
  {\rm sin}^2\left({\pi x/L}\right) }.
\label{gn0x}
\end{equation}
Note that $F_{\rm obc}(0,x) \sim x^{-2}$
for $|x|\ll \ell$. Therefore, the integral of $F_{\rm obc}(0,x)$ with respect to $x$ 
is infinite.

One may compare the results for OBC with those with PBC, for which
\begin{equation}
  G_{\rm pbc}(\vec{r}_1,\vec{r}_2) =  
  {(\pi/L)^{1/4} \over |\sin\pi(Z_-+iY)|^{1/4}}.
  \label{gcftpbcis}
\end{equation}
Again, setting $Y=0$, one obtains the two-point function $G(x,y)$ at the
critical point, from which we obtain $R_\xi^* \approx 0.187790$ and
$R_G^*(1/8,1/4)\approx 1.165899$.


\begin{thebibliography}{99}

\bibitem{Zurek-82} W. H. Zurek, Environment-induced superselection
  rules, Phys. Rev. D {\bf 26}, 1862 (1982).

\bibitem{Zurek-03}
  W. H. Zurek, Decoherence, einselection, and the quantum origins
  of the classical, Rev. Mod. Phys. {\bf 75}, 715 (2003).

\bibitem{Gaudin-76}
  M. Gaudin, Diagonalisation d'une classe d'Hamiltoniens de spin,
  J. Phys. (Paris) {\bf 37}, 1087 (1976).

\bibitem{PS-00}
  N. V. Prokof'ev and P. C. E. Stamp,
  Theory of the spin bath, Rep. Prog. Phys. {\bf 63}, 669 (2000).
  
\bibitem{CPZ-05} F. M. Cucchietti, J. P. Paz, and W. H. Zurek,
  Decoherence from spin environments, Phys. Rev. A {\bf 72}, 052113
  (2005).

\bibitem{QSLZS-06} H. T. Quan, Z. Song, X. F. Liu, P. Zanardi, and
  C. P. Sun, Decay of Loschmidt echo enhanced by quantum criticality,
  Phys. Rev. Lett. {\bf 96}, 140604 (2006).

\bibitem{RCGMF-07} D. Rossini, T. Calarco, V. Giovannetti,
  S. Montangero, and R. Fazio, Decoherence induced by interacting
  quantum spin baths, Phys. Rev. A {\bf 75}, 032333 (2007).

\bibitem{CFP-07} F. M. Cucchietti, S. Fernandez-Vidal, and J. P. Paz,
  Universal decoherence induced by an environmental quantum phase
  transition, Phys. Rev. A {\bf 75}, 032333 (2007).

\bibitem{YZL-07}
  Z.-G. Yuan, P. Zhang, and S.-S. Li, Loschmidt echo and Berry phase
  of a quantum system coupled to an $XY$ spin chain: Proximity
  to a quantum phase transition, Phys. Rev. A 75, 012102 (2007).

\bibitem{CP-08} C. Cormick and J. P. Paz, Decoherence induced by a
  dynamic spin environment: The universal regime,
  Phys. Rev. A {\bf 77}, 022317 (2008).
  
\bibitem{Zurek-09} W. H. Zurek, Quantum Darwinism, Nat. Phys. {\bf 5},
  181 (2009).

\bibitem{BESSS-10}
  M. Bortz, S. Eggert, C. Schneider, R. St{\"u}bner, and J. Stolze,
  Dynamics and decoherence in the central spin model using exact methods,
  Phys. Rev. B {\bf 82}, 161308(R) (2010).
  
\bibitem{DQZ-11} B. Damski, H. T. Quan, and W. H. Zurek, Critical
  dynamics of decoherence, Phys. Rev. A {\bf 83}, 062104 (2011).

\bibitem{WL-12}
  B.-B. Wei and R.-B. Liu, Lee-Yang zeros and critical times in decoherence
  of a probe spin coupled to a bath, Phys. Rev. Lett. {\bf 109}, 185701 (2012).
  
\bibitem{NDD-12} T. Nag, U. Divakaran, and A. Dutta, Scaling of the
  decoherence factor of a qubit coupled to a spin chain driven across
  quantum critical points, Phys. Rev. B {\bf 86}, 020401(R) (2012).

\bibitem{SND-16}
  S. Suzuki, T. Nag, and A. Dutta,
  Dynamics of decoherence: Universal scaling of the decoherence factor,
  Phys. Rev. A {\bf 93}, 012112 (2016).

\bibitem{JH-17}
  R. Jafari, and H. Johannesson,  Decoherence from spin environments:
  Loschmidt echo and quasiparticle excitations,
  Phys. Rev. B {\bf 96}, 224302 (2017).

\bibitem{V-18} E. Vicari, Decoherence dynamics of qubits coupled to
  systems at quantum transitions, Phys. Rev. A {\bf 98}, 052127
  (2018).

\bibitem{FCV-19}
  E. Fiorelli, A. Cuccoli, and P. Verrucchi,
  Critical slowing down and entanglement protection,
  Phys. Rev. A {\bf 100}, 032123 (2019).
  
\bibitem{RV-19} D. Rossini and E. Vicari, Scaling of decoherence and
  energy flow in interacting quantum many-body systems, Phys. Rev. A
  {\bf 99}, 052113 (2019).

\bibitem{HGMPM-19}
  P. Haikka, J. Goold, S. McEndoo, F. Plastina, and S. Maniscalco,
  Non-Markovianity, Loschmidt echo, and criticality: A unified picture,
  Phys. Rev. A {\bf 85}, 060101(R) (2012).
  
\bibitem{LZZ-19}
  F. Liu, X. Zhou, and Z.-W. Zhou,
  Memory effect and non-Markovian dynamics in an open quantum system,
  Phys. Rev. A {\bf 99}, 052119 (2019).
  
\bibitem{LSSWY-21}
  J.-X. Liu, H.-L. Shi, Y.-H. Shi, X.-H. Wang, and W.-L. Yang,
  Entanglement and work extraction in the central-spin quantum battery,
  Phys. Rev. B {\bf 104}, 245418 (2021).

\bibitem{YZWS-25}
  H.-Y. Yang, K. Zhang, X.-H. Wang, and H.-L. Shi,
  Optimal energy storage and collective charging speedup in the
  central-spin quantum battery, Phys. Rev. B {\bf 111}, 085410 (2025).

\bibitem{LHMC-08}
  C.-Y. Lai, J.-T. Hung, C.-Y. Mou, and P. Chen,  Induced decoherence
  and entanglement by interacting quantum spin baths,
  Phys. Rev. B {\bf 77}, 205419 (2008).

\bibitem{VPM-15}
  R. Vasseur, S. A. Parameswaran, and J. E. Moore,
  Quantum revivals and many-body localization,
  Phys. Rev. B {\bf 91}, 140202(R) (2015).

\bibitem{YZ-20}
  C.-Z. Yao and W.-M. Zhang,  Probing topological states through the exact
  non-Markovian decoherence dynamics of a spin coupled to a spin bath
  in the real-time domain, Phys. Rev. B {\bf 102}, 035133 (2020).

\bibitem{FRV-22} A.  Franchi, D. Rossini, and E. Vicari, Quantum
  many-body spin rings coupled to ancillary spins: The sunburst
  quantum Ising model, Phys. Rev. E {\bf 105}, 054111 (2022).

\bibitem{FRV-22-2} A.  Franchi, D. Rossini, and E. Vicari, Decoherence
  and energy flow in the sunburst quantum Ising model,
  J.~Stat.~Mech. (2022) 083103.  

\bibitem{MS-23} A. Mitra and S. C. L. Srivastava, Sunburst quantum
  Ising model under interaction quench: entanglement and role of
  initial state coherence, Phys. Rev. E {\bf 108}, 054114 (2023).

\bibitem{FT-23} A. Franchi and F. Tarantelli, Liouvillian gap and
  out-of-equilibrium dynamics of a sunburst Kitaev ring: from local
  to uniform dissipation, Phys. Rev. B {\bf 108}, 094114 (2023).
  
\bibitem{MS-24}
  A. Mitra and S. C. L. Srivastava,  Sunburst quantum Ising battery,
  Phys. Rev. A {\bf 110}, 012227 (2024).

  
\bibitem{BYZB-20}
  S. Barbarino, J. Yu, P. Zoller, and J. C. Budich,
  Preparing atomic topological quantum matter by adiabatic nonunitary dynamics,
  Phys. Rev. Lett. {\bf 124}, 010401 (2020).

\bibitem{BBSD-21}
  S. Bhattacharjee, S. Bandyopadhyay, D. Sen, and A. Dutta,
  Bilayer Haldane system: Topological characterization and
  adiabatic passages connecting Chern phases,
  Phys. Rev. B {\bf 103}, 224304 (2021).

\bibitem{MFPH-23}
  M. Manna{\"i}, J.-N. Fuchs, F. Pi{\'e}chon, and S. Haddad,
  Stacking-induced Chern insulator, Phys. Rev. B {\bf 107}, 045117 (2023).
  
\bibitem{FPV-23} A. Franchi, A. Pelissetto, and E. Vicari, Quantum
  critical behaviors and decoherence of weakly coupled quantum Ising
  models within an isolated global system,
  Phys. Rev. E {\bf 107}, 014113 (2023).  

\bibitem{SGCS-97} S. L. Sondhi, S. M. Girvin, J. P. Carini, and
  D. Shahar, Continuous quantum phase transitions,
  Rev. Mod. Phys. {\bf 69}, 315 (1997).   

\bibitem{Sachdev-book} S. Sachdev, {\it Quantum Phase Transitions}, 2nd ed.
  (Cambridge University Press, Cambridge, 2011).

\bibitem{RV-21} D. Rossini and E. Vicari, Coherent and dissipative
  dynamics at quantum phase transitions, Phys. Rep. {\bf 936}, 1 (2021).

\bibitem{Schollwock-05}
  U. Schollw{\"o}ck, The density-matrix renormalization group,
  Rev. Mod. Phys. {\bf 77}, 259 (2005).
  
\bibitem{KNK-81} M. Kohmoto, M. den Nijs, and L. P. Kadanoﬀ,
  Hamiltonian studies of the $d=2$ Ashkin-Teller model, Phys. Rev. B
  {\bf 24}, 5229 (1981).

\bibitem{IS-84} F. Igloi and J. Solyom, Phase diagram and critical
  properties of the (1+1)-dimensional Ashkin-Teller model, J. Phys. A
  {\bf 17}, 1531 (1984).

\bibitem{AF-84} F. C. Alcaraz and J. R. Drugowich de Felicio, Finite
  size studies of the Ashkin-Teller model, J. Phys. A {\bf 17} L651 (1984).

\bibitem{Fradkin-84}    
  E. Fradkin, $N$-Color Ashkin-Teller model in two dimensions:
  Solution in the large-$N$ limit, Phys. Rev. Lett. {\bf 53}, 1967 (1984).

\bibitem{Shankar-85}
  R. Shankar, Ashkin-Teller and Gross-Neveu models: New relations and results,
  Phys. Rev. Lett. {\bf 55}, 453 (1985).

\bibitem{GR-87} G. von Gehlen and V. Rittenberg, The Ashkin-Teller
  quantum chain and conformal invariance, J. Phys. A {\bf 20}, 227 (1987).

\bibitem{ABB-88} F. C. Alcaraz, M. N. Barber, and M. T. Batchelor,
  Finite size studies of the Ashkin-Teller model,
  Ann. Phys. {\bf 182}, 280 (1988).

\bibitem{AT-43} J. Ashkin and E. Teller, Statistics of Two-Dimensional
  Lattices with Four Components, Phys. Rev. {\bf 64}, 178 (1943).

\bibitem{WL-74}
  F. Y. Wu and K. Y. Liu,
  Two phase transitions in the Ashkin-Teller model,
  J. Phys. C: Solid State Phys. {\bf 7}, L181 (1974).

\bibitem{DACDRS-book}
  A. Dutta, G. Aeppli, B. K. Chakrabarti,
  U. Divakaran, T. F. Rosenbaum, and D. Sen, {\em Quantum phase
    transitions in transverse field spin models: From statistical
    physics to quantum information}, (Cambridge University Press, 2015).

\bibitem{Pfeuty-70}
  P. Pfeuty,
  The one-dimensional Ising model with a transverse field,
  Ann. Phys. {\bf 57}, 79 (1970).
  
\bibitem{CPV-15}
  M. Campostrini, A. Pelissetto, and E. Vicari,
  Quantum Ising chains with boundary fields,
  J. Stat. Mech. P11015 (2015).

\bibitem{CPV-14} M. Campostrini, A. Pelissetto, and E. Vicari,
  Finite-size scaling at quantum transitions, Phys. Rev. B {\bf 89},
  094516 (2014).

\bibitem{PV-02} A. Pelissetto and E. Vicari, 
  Critical phenomena and renormalization group theory, 
  Phys. Rep. {\bf 368}, 549 (2002).

\bibitem{CHPV-02}  
  M. Caselle, M. Hasenbusch, A. Pelissetto, and E. Vicari, 
  Irrelevant operators in the two-dimensional Ising model, 
  J. Phys. A {\bf 35}, 4861 (2002).

\bibitem{BGR-87} M. Baake, G. von Gehlen and V. Rittenberg, Operator
  content of the Ashkin-Teller quantum chain-superconformal and
  Zamolodchikov-Fateev invariance: I. Free boundary conditions,
  J. Phys {\bf 20}, L479 (1987); Operator content of
  the Ashkin-Teller quantum chain, superconformal and
  Zamolodchikov-Fateev invariance. II. Boundary conditions compatible
  with the torus, J. Phys. A {\bf 20}, L487 (1987).

\bibitem{Yang-87} S. Yang, Modular invariant partition function of the
  Ashkin-Teller model on the critical line and N = 2 superconformal
  invariance, Nucl. Phys. B {\bf 285}, 183 (1987).

\bibitem{YZ-87} S. Yang and H. Zheng, Superconformal invariance in
  the two-dimensional Ashkin-Teller model, Nuc. Phys. B {\bf 285}, 410
  (1987).

\bibitem{YHK-94}
  M. Yamanaka, Y. Hatsugai, and M. Kohmoto,
  Phase diagram of the Ashkin-Teller quantum spin chain,
  Phys. Rev. B {\bf 50}, 559 (1994).
  
\bibitem{YK-95}
  M. Yamanaka and M. Kohmoto,
  Line of continuously varying criticality in the Ashkin-Teller quantum chain,
  Phys. Rev. B {\bf 52}, 1138 (1995).

\bibitem{BBBD-15} J. C. Bridgeman, A O'Brien, S. D. Barlett, and
  A. C. Doherty, Multiscale entanglement renormalization ansatz for
  spin chains with continuously varying criticality,
  Phys. Rev. B {\bf 91}, 165129 (2015).

\bibitem{BBDF-15} A O'Brien, S. D. Barlett, A. C. Doherty, and
  S. T. Flammia, Symmetry-respecting real-space renormalization for
  the quantum Ashkin-Teller model, Phys. Rev. E {\bf 92}, 042163 (2015).

\bibitem{LMC-24} B. E. L\"uscher, F. Mila, and N. Chepiga, Critical
  properties of the quantum Ashkin-Teller chain with chiral
  perturbations, Phys. Rev. B {\bf 108}, 184425 (2023)

\bibitem{LCFO-26}
  Y. Liu, N. Chepiga, Y. Fukusumi, and M. Oshikawa,
  Boundary critical phenomena in the quantum Ashkin-Teller model,
  arXiv:2601.16951.
  
\bibitem{Wu-77}
  F. Y. Wu,
  Ashkin-Teller Model as a Vertex Problem,
  J. Math. Phys. {\bf 18}, 611 (1977).
  
\bibitem{Baxter-82}
  R. J. Baxter, {\em Exactly solvable models in statistical mechanics},
  Academic Press, New York (1982).

\bibitem{DR-79}
  E. Domany and E. K. Riedel,
  Two-dimensional anisotropic $N$-vector models,
  Phys. Rev. B {\bf 19}, 5817 (1979).

\bibitem{DK-82}
  J. R. Drugowich de Fel\'icio and R. K\"oberle, 
  Critical exponents of the Ashkin-Teller model,
  Phys. Rev. B {\bf 25}, 511 (1982).

\bibitem{WD-93} S. Wiseman and E. Domany, A Cluster Method for the
  Ashkin–Teller Model, Phys. Rev. E {\bf 48}, 4080 (1993)

\bibitem{KKD-97}
  G. Kamieniarz, P. Kozlowski, and R. Dekeyser,
  Critical Ising lines of the $d=2$ Ashkin-Teller model,
  Phys. Rev. E {\bf 55},  3724 (1997). 

\bibitem{DG-04}
  G. Delfino and P. Grinza,
  Universal ratios along a line of critical points.
  The Ashkin-Teller model,  Nucl. Phys. B {\bf 682}, 521 (2004).
  
\bibitem{GM-05}
  A. Giuliani and V. Mastropietro,
  Anomalous universality in the anisotropic Ashkin-Teller model,
  Commun. Math. Phys. {\bf 256}, 681 (2005).
  
\bibitem{MM-23} I. Mukherjee and P. K. Mohanty, Hidden
  superuniversality in systems with continuous variation of critical
  exponents, Phys. Rev. B {\bf 108}, 174417 (2023).

\bibitem{KB-23} I. Kecoglu and A. N. Berker, Global Ashkin-Teller
  Phase Diagrams in Two and Three Dimensions: Multicritical
  Bifurcation versus Double Tricriticality - Endpoint,
  Physica A {\bf 630}, 129248 (2023).

\bibitem{ADG-24}
  Y. Aoun, M. Dober, and A. Glazman,
  Phase diagram of the Ashkin-Teller model,
  Commun. Math. Phys. 4{\bf 05}, 37 (2024).

\bibitem{Dober-25} M. Dober, On antiferromagnetic regimes in the
  Ashkin-Teller model, Electron. J. Probab. {\bf 30}, 1 (2025).

\bibitem{It-Dr-book} C. Itzykson and J. M. Drouffe,
  {\em Statistical Field Theory} (Cambridge Univ. Press, Cambridge 1989).

\bibitem{CFT-book}
  P. Di Francesco, P. Mathieu, and D. Senechal,
  {\em Conformal Field Theory} (Springer Verlag, New York, 1997).

\bibitem{Khan-etal-17} N. Khan, P. Sarkar, A. Midya,
  P. Mandal, and P. K. Mohanty, Continuously Varying Critical Exponents
  Beyond Weak Universality, Sci. Rep. {\bf 7}, 45004 (2017).

\bibitem{Kadanoff-77} L. P. Kadanoff, Connections between the critical
  behavior of the planar model and that of the eight-vertex model,
  Phys. Rev. Lett. {\bf 39}, 903 (1977); Multicritical behavior at the
  Kosterlitz-Thouless critical point, Ann. Phys. {\bf 120}, 39 (1979).

\bibitem{AFOV-08}
  L. Amico, R. Fazio, A. Osterloh, and V. Vedral,
  Entanglement in many-body systems,
  Rev. Mod. Phys. {\bf 80}, 517 (2008).

\bibitem{Gu-10}
  S.-J. Gu, Fidelity approach to quantum phase transitions,
  Int. J. Mod. Phys. B {\bf 24}, 437 (2010).

\bibitem{BAB-17} D. Braun, G. Adesso, F. Benatti, R. Floreanini,
  U. Marzolino, M. W. Mitchell, and S. Pirandola, Quantum-enhanced
  measurements without entanglement, 
  Rev. Mod. Phys. {\bf 90}, 035006 (2018).

\bibitem{RV-18} D. Rossini and E. Vicari, Ground-state fidelity at
  first-order quantum transitions, Phys. Rev. E {\bf 98}, 062137 (2018).

\bibitem{GZ-98} R. Guida and J. Zinn-Justin, Critical exponents of the
  $N$-vector model, J. Phys. A {\bf 31}, 8103 (1998).

\bibitem{CPRV-02} M. Campostrini, A. Pelissetto, P. Rossi, and
  E. Vicari, 25th order high-temperature expansion results for
  three-dimensional Ising-like systems on the simple cubic lattice,
  Phys. Rev. E {\bf 65}, 066127 (2002).

\bibitem{Hasenbusch-10} M. Hasenbusch, Finite-size scaling study of
  lattice models in the three-dimensional Ising universality class,
  Phys. Rev. B {\bf 82}, 174433 (2010).
  
\bibitem{KPSV-16} F. Kos, D. Poland, D. Simmons-Duffin, and A. Vichi,
  Precision islands in the Ising and O($N$) models,
  J. High Energ. Phys. 08 (2016) 036.  

\bibitem{KP-17} M. V. Kompaniets and E. Panzer, Minimally subtracted
  six-loop renormalization of $\phi^4$-symmetric theory and critical
  exponents, Phys. Rev. D {\bf 96}, 036016 (2017).
  
\bibitem{FXL-18} A. M. Ferrenberg, J. Xu, and D. P. Landau, Pushing
  the limits of Monte Carlo simulations for the three-dimensional
  Ising model, Phys. Rev. E {\bf 97}, 043301 (2018).  
  
\bibitem{Hasenbusch-21}
  M. Hasenbusch, Restoring isotropy in a three-dimensional lattice model:
  The Ising universality class, Phys. Rev. B {\bf 104}, 014426 (2021).

\bibitem{LF-72} K.-S. Liu and M. E. Fisher, Quantum lattice gas and
  the existence of a supersolid, J. Low Temp. Phys. {\bf 10}, 655
  (1972).

\bibitem{FN-74} M. E. Fisher and D. R. Nelson, Spin flop, supersolids,
  and bicritical and tetracritical points, Phys. Rev. Lett. {\bf 32},
  1350 (1974).

\bibitem{NKF-74} D. R. Nelson, J. M. Kosterlitz, and M. E. Fisher,
  Renormalization-group analysis of bicritical and tetracritical
  points, Phys. Rev. Lett.  {\bf 33}, 813 (1974); J. M. Kosterlitz,
  D. R. Nelson, and M. E. Fisher, Bicritical and tetracritical points
  in anisotropic antiferromagnetic systems, Phys. Rev. B {\bf 13}, 412
  (1976).

\bibitem{CPV-03} P. Calabrese, A. Pelissetto, and E. Vicari,
  Multicritical behavior of ${\rm O}(n_1)\oplus {\rm O}(n_2)$-symmetric
  systems, Phys. Rev. B {\bf 67}, 054505 (2003).

\bibitem{HV-11} M. Hasenbusch and E. Vicari, Anisotropic perturbations
  in 3D O(N) vector models, Phys. Rev. B {\bf 84}, 125136 (2011).

\bibitem{BPV-22-z2g} C. Bonati, A. Pelissetto, and E. Vicari,
  Multicritical point of the three-dimensional ${\mathbb Z}_2$ gauge
  Higgs model, Phys. Rev. B {\bf 105}, 165138 (2022).

\bibitem{CHPV-06} M. Campostrini, M. Hasenbusch, A. Pelissetto, and
  E. Vicari, Theoretical estimates of the critical exponents of the
  superfluid transition in $^4$He by lattice methods,
  Phys. Rev. B {\bf 74}, 144506 (2006).

\bibitem{CLLPSSV-20} S. M. Chester, W. Landry, J. Liu, D. Poland,
  D. Simmons-Duffin, N. Su, and A. Vichi, Carving out OPE space and
  precise O(2) model critical exponents, J. High Energ. Phys. {\bf 06},
  142 (2020).

\bibitem{BPV-25} C. Bonati, A. Pelissetto, and E. Vicari,
  Three-dimensional Abelian and non-Abelian gauge Higgs theories,
  Phys. Rep. {\bf 1133}, 1 (2025).

\bibitem{Hasenbusch-25} M. Hasenbusch, Eliminating leading and
  subleading corrections to scaling in the three-dimensional XY
  universality class, Phys. Rev. B {\bf 112}, 184512 (2025).
  
\bibitem{Hasenbusch-25-2} M. Hasenbusch, Precision estimates of large
  charge RG exponents $Y_q$  in the 3D XY universality class,
  arXiv:2511.18321.

\bibitem{CPV-00} J. Carmona, A. Pelissetto, and E. Vicari, The
  $N$-component Ginzburg-Landau Hamiltonian with cubic symmetry: a
  six-loop study, Phys. Rev. B {\bf 61}, 15136 (2000).

\bibitem{Weiss-book} U. Weiss, {\em Quantum dissipative systems}
  (World Scientific, Singapore, 2012).

\bibitem{CL-83} A. O. Caldeira and A. J. Leggett, Quantum tunnelling
  in a dissipative system, Ann. Phys. {\bf 149}, 374 (1983).

\bibitem{LCD-87} A. J. Leggett, S. Chakravarty, A. T. Dorsey,
  M. P. A. Fisher, A. Garg, and W. Zwerger, Dynamics of the dissipative
  two-state system, Rev. Mod. Phys. {\bf 59}, 1 (1987);
  Rev. Mod. Phys. {\bf 67}, 725 (Erratum).

\bibitem{WTS-04} P. Werner, M. Troyer, and S. Sachdev, Quantum spin
   chains with site dissipation, J. Phys. Soc. Jpn. Suppl. {\bf 74},
   67 (2005).

\bibitem{SWT-04} S. Sachdev, P. Werner, and M. Troyer, Universal
  conductance of nanowires near the superconductor-metal quantum
  transition, Phys. Rev. Lett. {\bf 92}, 237003 (2004).
     
\bibitem{WVTC-05} P. Werner, K. V\"olker, M. Troyer, and
  S. Chakravarty, Phase diagram and critical exponents of a
  dissipative Ising spin chain in a transverse magnetic field,
  Phys. Rev. Lett. {\bf 94}, 047201 (2005).

\bibitem{YMZ-14} S. Yin, P. Mai, and F. Zhong, Nonequilibrium quantum
  criticality in open systems: The dissipation rate as an additional
  indispensable scaling variable, Phys. Rev. B {\bf 89}, 094108
  (2014).
  
\bibitem{ARBA-17} O. Alberton, J. Ruhman, E. Berg, and E. Altman, Fate
  of the One Dimensional Ising Quantum Critical Point Coupled to a
  Gapless Boson, Phys. Rev. B {\bf 95}, 075132 (2017).
  
\bibitem{KMSFR-17}
  M. Keck, S. Montangero, G. E. Santoro, R. Fazio, and D. Rossini,
  Dissipation in adiabatic quantum computers: lessons from an exactly
  solvable model, New. J. Phys. {\bf 19}, 113029 (2017).

\bibitem{WSBR-18} H. Weisbrich, C. Saussol, W. Belzig, and
  G. Rastelli, Decoherence in the quantum Ising model with transverse
  dissipative interaction in the strong coupling regime, Phys. Rev. A
  {\bf 98}, 052109 (2018).

\bibitem{NRV-19} D. Nigro, D. Rossini, and E. Vicari, Competing
  coherent and dissipative dynamics close to quantum criticality,
  Phys. Rev. A {\bf 100}, 052108 (2019); D. Rossini and E. Vicari,
  Scaling behavior of stationary states arising from dissipation at
  continuous quantum transitions, Phys. Rev. B {\bf 100}, 174303
  (2019).

\bibitem{RV-20} D. Rossini and E. Vicari, Dynamic Kibble-Zurek scaling
  framework for open dissipative many-body systems crossing quantum
  transitions, Phys. Rev. Res. {\bf 2}, 023211 (2020).

\bibitem{BP-book} H.-P. Breuer and F. Petruccione, {\em The Theory of
  Open Quantum Systems} (Oxford University Press, New York, 2002).

\bibitem{RH-book} A. Rivas and S. F. Huelga, {\em Open Quantum Systems:
  An Introduction} (SpringerBriefs in Physics, Springer, 2012).

\bibitem{LCF-18}
  Y. Li, X. Chen, and M. P. A. Fisher, Quantum Zeno effect and the
  many-body entanglement transition, Phys. Rev. B {\bf 98}, 205136 (2018).

\bibitem{LCF-19}
  Y. Li, X. Chen, and M. P. A. Fisher, Measurement-driven entanglement
  transition in hybrid quantum circuits, Phys. Rev. B {\bf 100}, 134306 (2019).
  
\bibitem{CTD-19}
  X. Cao, A. Tilloy, and A. De Luca, Entanglement in a fermion chain
  under continuous monitoring, SciPost Phys. {\bf 7}, 24 (2019).
  
\bibitem{SRN-19}
  B. Skinner, J. Ruhman, and A. Nahum, Measurement-induced phase transitions
  in the dynamics of entanglement, Phys. Rev. X {\bf 9}, 031009 (2019).

\bibitem{RV-20B}
  D. Rossini and E. Vicari, Measurement-induced dynamics of many-body
  systems at quantum criticality, Phys. Rev. B {\bf 102}, 035119 (2020).
  
\bibitem{TPDC-24}
  M. Tsitsishvili, D. Poletti, M. Dalmonte, and G. Chiriac{\`o},
  Measurement induced transitions in non-Markovian free fermion ladders,
  SciPost Phys. Core {\bf 7}, 011 (2024).

\bibitem{MTC-25}
  C. Muzzi, M. Tsitsishvili, and G. Chiriac{\`o},
  Entanglement enhancement induced by noise in inhomogeneously
  monitored systems Phys. Rev. B {\bf 111}, 014312 (2025).

\bibitem{CV-10} M. Campostrini and E. Vicari, Trap-size scaling in
  confined-particle systems at quantum transitions,
  Phys. Rev. A {\bf 81}, 023606 (2010).

\end{thebibliography}
\end{document}